\documentclass[a4paper,11pt]{article}
\usepackage{jheppub}

\usepackage[whole]{bxcjkjatype} 
\usepackage{physics}
\usepackage{braket}

\usepackage{graphicx}
\usepackage{tikz}
\usepackage{tikz-cd}
\usepackage{tikzit}
\usepackage{mathtools}

\tikzstyle{pointoperator}=[fill=black, draw=none, shape=circle, minimum size=.1 cm, inner sep=0 pt]
\tikzstyle{red dot}=[fill=red, draw=red, shape=circle, minimum size=.1 cm, inner sep=0 pt]
\tikzstyle{red diamond}=[fill=red, draw=red, shape=diamond]
\tikzstyle{blue dot}=[fill=blue, draw=blue, shape=circle]

\tikzstyle{dashedline}=[dashed, -, thick]
\tikzstyle{blueline}=[-, draw=blue, thick]
\tikzstyle{arrowline}=[<-, thick]
\tikzstyle{normalline}=[-, thick, draw={rgb,255: red,171; green,171; blue,171}]
\tikzstyle{fillline}=[-, thick, fill=red, fill opacity=.5]
\tikzstyle{bluefillline}=[-, fill opacity=.5, fill=blue, thick]
\tikzstyle{purplefillline}=[-, fill opacity=.5, fill=purple, thick]
\tikzstyle{bluedashedline}=[-, draw=blue, thick, dashed]

\newcommand{\ntau}{n_{\tau}}
\newcommand{\nxy}{n_{xy}}
\newcommand{\Dtau}{\Delta_{\tau}}
\newcommand{\Dx}{\Delta_{x}}
\newcommand{\Dy}{\Delta_{y}}

\newcommand{\xhat}{\hat{x}}
\newcommand{\yhat}{\hat{y}}
\newcommand{\tauhat}{\hat{\tau}}
\newcommand{\DtauPhi}{\Dtau \phi + 2\pi \ntau}
\newcommand{\DxyPhi}{\Dx \Dy \phi + 2\pi \nxy}
\newcommand{\btheta}{\theta_{\mathrm{bul}}}
\newcommand{\ftheta}{\theta_{\mathrm{fol}}^{x}}
\newcommand{\sbulk}{S_{\mathrm{bulk}}}
\newcommand{\sfol}{S_{\mathrm{fol}}^{x}}

\title{Exotic theta terms in 2+1d fractonic field theory}

\preprint{UT-Komaba/26-3}
\author[a]{Yuki Furukawa}
\affiliation[a]{Graduate School of Arts and Sciences, University of Tokyo, Komaba, \\ Meguro-ku, Tokyo 153-8902, Japan}

\emailAdd{furukawa@hep1.c.u-tokyo.ac.jp}

\abstract{
In this work, we study exotic theta terms in the 2+1d $\phi$-theory, which provides a continuum description of the XY-plaquette model.
The $\phi$-theory can be viewed as a fractonic analogue of the 1+1d compact boson and exhibits momentum and winding subsystem symmetries.
In this theory, discontinuous field configurations play a crucial role.
Although such configurations spoil the naive topology of the field, they induce nontrivial backreactions that give rise to new topological terms.
We study two types of theta terms, which we call the bulk theta term and the foliated theta term.
The foliated theta term is constructed by coupling winding currents on neighboring leaves of a foliation.
Remarkably, the corresponding theta angle can vary spatially without affecting the classical equations of motion.
Both theta terms lead to generalized Witten effects, in which vortex operators carrying winding subsystem charge acquire momentum subsystem charge.
In the case of the foliated theta angle, the Witten effect exhibits a more intricate structure: vortex operators acquire a quadrupolar momentum charge.
We demonstrate these features using lattice realizations based on the modified Villain formulation.
}

\begin{document}
\maketitle
\flushbottom

\section{Introduction}

Exotic lattice models with subsystem symmetry have attracted considerable attention in recent years.
Subsystem symmetry is characterized by a set of symmetry operators acting on subsystems.
Unlike ordinary global symmetries, these operators are not fully topological; rather, they can be deformed only within the subsystem.

Subsystem symmetry plays an important role in the understanding of a class of gapped phases called fracton phases~\cite{Nandkishore:2018sel,Pretko:2020cko}.
Such phases host excitations with restricted mobility called fractons, which reflect the underlying subsystem symmetry.
Field theory descriptions of such phases have been developed~\cite{Slagle:2017wrc,Pretko:2017xar,Pretko:2018jbi,Slagle:2018swq,You:2019ciz,Seiberg:2020bhn,Seiberg:2020wsg,Seiberg:2020cxy,Gorantla:2020xap,Slagle:2020ugk,Gorantla:2020jpy,Yamaguchi:2021qrx,Hsin:2021mjn,Gorantla:2021bda,Geng:2021cmq,Burnell:2021reh,Yamaguchi:2021xeq,Gorantla:2022eem,Gorantla:2022ssr,Ohmori:2022rzz,Honda:2022shd,Hsin:2023ooo,Shimamura:2024kwf,Ebisu:2024mbb}.

There are also gapless lattice models with subsystem symmetry.
For example, the 2+1 dimensional XY-plaquette model~\cite{Paramekanti:2002iup} in an appropriate parameter region, which is a main focus of this work, is gapless and has subsystem symmetry.
In this model, discontinuous field configurations play a crucial role in the physics, leading to UV/IR mixing~\cite{Seiberg:2020bhn,Gorantla:2021bda}.
This system is described by a gapless fractonic field theory, known as the 2+1d $\phi$-theory~\cite{Seiberg:2020bhn},\footnote{Although gapless fracton phases also exist, the $\phi$-theory does not host fractonic excitations. Nevertheless, we refer to it as "fractonic" in the sense that subsystem symmetry plays a central role.} which is reminiscent of the standard compact boson.

On the other hand, topological terms play an important role in quantum field theory.
A topological term is determined entirely by the topological sector of a field configuration and therefore does not affect the classical equations of motion.
Some of these terms cause the Witten effect \cite{Witten:1979ey}, in which magnetic operators are dressed with fractional electric charge.
What happens to topological terms in fractonic field theories?
At first sight, discontinuous field configurations appear to obscure the topology of the fields, obstructing the existence of conventional topological terms.
However, as we will show, the opposite phenomenon can also occur: topological terms that are originally regarded as trivial, in the sense that they always vanish, can become nontrivial due to field discontinuities.
In this work, we investigate the properties of such exotic topological terms in the 2+1d $\phi$-theory and their lattice realization.

We briefly comment on previous works:
\begin{itemize}
    \item In \cite{Pretko:2017xar}, Pretko studied theta terms and the Witten effects in 3+1d tensor gauge theories in the continuum.
    \item In \cite{Bedogna:2026bck}, Bedogna and Mancani introduced a theta term in the 2+1d $\phi$-theory, which we refer to as the bulk theta term.
    We further analyze its properties, including the Witten effect.
    In particular, we provide a lattice realization and highlight the subtlety in the periodicity of the theta angle.
\end{itemize}

The rest of this paper is organized as follows.
In Section~\ref{section_review}, we review the $\phi$-theory both on the lattice and in the continuum, with particular emphasis on a lattice realization known as the modified Villain formulation.
In Section~\ref{section_lattice}, we study the lattice realization of the theta terms.
In Section~\ref{section_continuum}, we then discuss their continuum description.
In Section~\ref{section_conclusion}, we summarize our results and comment on future directions.
In Appendix~\ref{Appendix_topological_charge}, we construct the topological charge~(\ref{cont_topological_charge}) in the original XY-plaquette model without using the modified Villain formulation.
In Appendix~\ref{appendix_quantization_of_topological_charge}, we show that the topological charge~(\ref{cont_topological_charge}) is quantized to integer values in the continuum $\phi$-theory.

\section{Review of the $\phi$-theory}
\label{section_review}
In this section, we review the 2+1d $\phi$-theory, which is known as a continuum description of the XY-plaquette model.
This theory has been studied in \cite{Seiberg:2020bhn,Gorantla:2021bda,Burnell:2021reh,Spieler:2024fby,Ohmori:2025fuy,Apruzzi:2025mdl,Bedogna:2026bck}.
We also review the lattice construction of the theory proposed in \cite{Gorantla:2021svj}, using a technique called the modified Villain formulation.
Throughout this paper, spacetime is assumed to be a three-dimensional torus.

\subsection{Continuum theory}
\label{review_of_continuum}

We begin with the 2+1d XY-plaquette model~\cite{Paramekanti:2002iup}.
At each site $n$ on the two-dimensional square lattice, we assign a field $\phi(n)$ with periodicity $\phi(n)\sim\phi(n)+2\pi$.
We impose the commutation relations $[\phi(n),\pi(n')]=i\delta_{n,n'},\,[\phi(n),\phi(n')]=[\pi(n),\pi(n')]=0$,
where $\pi(n)$ is the canonical momentum conjugate to $\phi(n)$.
The Hamiltonian is given by
\begin{align}
    \label{original_hamiltonian}
    H = \sum_{n} \frac{U}{2} \pi^2(n) - \sum_{n} K \cos(\Delta_x \Delta_y \phi(n)),
\end{align}
where $\Delta_{\mu}$ denotes the lattice derivative, and $U$ and $K$ are positive real parameters.
For example, $\Delta_x \Delta_y \phi(n) = \phi(n+\hat{x}+\hat{y}) - \phi(n+\hat{x}) - \phi(n+\hat{y}) + \phi(n)$,
where $\hat{\mu}$ is the unit vector in the $\mu$-direction.

If $K \gg U$, $\Delta_x \Delta_y \phi(n)$ is energetically constrained to be close to $0 \ \mathrm{mod}\, 2\pi$,
leading to the Euclidean action of a continuum field theory known as the $\phi$-theory~\cite{Seiberg:2020bhn}:
\begin{align}
    \label{cont_action}
    S_0 = \int d\tau \, dx \, dy\, \left\{\frac{\mu_0}{2}(\partial_{\tau}\phi)^2 + \frac{1}{2\mu} (\partial_{x}\partial_{y}\phi)^2\right\}.
\end{align}
Here, $\phi$ is a periodic scalar field with a non-standard identification $\phi \sim \phi + 2\pi m^x(x)+ 2\pi m^y(y)$,
where $m^x(x)$ and $m^y(y)$ are integer-valued piecewise constant functions.
A more precise definition will be given below.
The equation of motion reads
\begin{align}
    \label{cont_eom}
    \mu_0 \partial_{\tau}^2 \phi = \frac{1}{\mu}\partial^2_{x}\partial^2_{y}\phi.
\end{align}

It has been shown that discontinuous field configurations play an important role in this theory~\cite{Seiberg:2020bhn,Gorantla:2021bda}.
One way to see this is to note that a discontinuous configuration $\phi = \Theta(x)$\footnote{Here we ignore boundary conditions.} is a zero-energy solution of the equation of motion~(\ref{cont_eom}),
where $\Theta(x)$ is the step function
\begin{align}
    \Theta(x) =
    \begin{cases}
        1 & x > 0,\\
        0 & x < 0.
    \end{cases}
\end{align}
In terms of the original XY-plaquette model, $\Delta_x \Delta_y \phi \ \mathrm{mod}\, 2\pi$ is energetically suppressed when $K \gg U$,
whereas $\Delta_x \phi$ is not.

Let us comment on deformations of the minimal action~(\ref{cont_action}).
One can add higher-derivative (or higher-order) terms to the action without breaking the momentum and/or winding subsystem symmetries discussed below.
As shown in \cite{Seiberg:2020bhn}, because discontinuous field configurations are not suppressed in this theory,
higher-derivative terms can quantitatively modify the spectrum of modes charged under the momentum and/or winding symmetries even in the continuum limit,
while the qualitative scaling $\frac{1}{a}$, where $a$ is the lattice spacing, is preserved.

\paragraph{Global definition of $\phi$}
A field configuration of $\phi$ is defined in a manner similar to that of the standard compact scalar field~\cite{Seiberg:2020bhn}.
First, we take an open cover of the spacetime manifold (i.e., $T^3$ in our case).
On each patch $U_{\alpha}$, $\phi$ is locally described by a real-valued function $\phi_{\alpha}$.
On the overlap $U_{\alpha} \cap U_{\beta}$ of two patches, a transition function $m_{\alpha\beta}(x,y)$ relates the two local expressions of $\phi$ as
\begin{align}
    \phi_{\alpha}(\tau,x,y) = \phi_{\beta}(\tau,x,y) + 2\pi m_{\alpha\beta}(x,y).
\end{align}
We assume that $m_{\alpha\beta}$ takes the form
\begin{align}
    m_{\alpha\beta}(x,y) = m^{x}_{\alpha\beta}(x) + m^{y}_{\alpha\beta}(y),
\end{align}
where $m^{x}_{\alpha\beta}(x)$ and $m^{y}_{\alpha\beta}(y)$ are integer-valued (possibly discontinuous) functions
depending only on the $x$- and $y$-coordinates, respectively.
Accordingly, $m_{\alpha\beta}$ satisfies the cocycle conditions
\begin{align}
    m_{\alpha\beta} + m_{\beta\gamma} + m_{\gamma\alpha} = 0,\\
    m_{\beta\alpha} = - m_{\alpha\beta}.
\end{align}
A configuration of $\phi$ is specified by the data of an open cover $\{U_{\alpha}\}_{\alpha}$,
a collection of local functions $\{\phi_{\alpha}\}_{\alpha}$, and a collection of transition functions
$\{m_{\alpha\beta}\}_{\alpha,\beta}$ defined on each overlap.

Physically, this description contains redundancies, which should be regarded as gauge redundancies.
One such redundancy is the choice of the open cover.
For example, one can refine the cover by introducing an additional patch $U_{\alpha}$ and a corresponding $\phi_{\alpha}$ in a consistent way.
We require that the physics be independent of this choice.
Another redundancy lies in the choice of $\{\phi_{\alpha}\}_{\alpha}$ and $\{m_{\alpha\beta}\}_{\alpha,\beta}$.
Specifically, for integer-valued functions $k^x_{\alpha}(x)$ and $k^y_{\alpha}(y)$,
\begin{align}
    \phi'_{\alpha}(\tau,x,y) &= \phi_{\alpha}(\tau,x,y) + 2\pi k^x_{\alpha}(x) + 2\pi k^y_{\alpha}(y),\\
    m^{'x}_{\alpha\beta}(x) &= m^x_{\alpha\beta}(x) + k^x_{\alpha}(x) - k^x_{\beta}(x),\\
    m^{'y}_{\alpha\beta}(y) &= m^y_{\alpha\beta}(y) + k^y_{\alpha}(y) - k^y_{\beta}(y)
\end{align}
describe the same physical configuration.

In summary, once an open cover, local functions, and transition functions are specified, they determine a global configuration of $\phi$.
However, this description is redundant, and the theory must be invariant under the corresponding gauge transformations.
The operators $e^{i\phi}$, $\partial_{\tau}\phi$, and $\partial_{x}\partial_{y}\phi$ are gauge invariant,
whereas $\partial_{x}\phi$ and $\partial_{y}\phi$ are not well-defined.

Next, we review two types of subsystem symmetries of the $\phi$-theory.
\paragraph{Momentum symmetry}
The action~(\ref{cont_action}) is invariant under
\begin{align}
    \label{cont_2+1d_phi_theory_momentum_trsf}
    \phi(\tau,x,y) \to \phi(\tau,x,y) + c_x(x) + c_y(y),
\end{align}
where $c_x$ and $c_y$ are real-valued functions that depend only on the $x$- and $y$-coordinates, respectively.
Due to the gauge redundancies explained above,
the parameters of the transformation~(\ref{cont_2+1d_phi_theory_momentum_trsf}) are subject to the identifications
$c_x(x) \sim c_x(x) + 2\pi m_x(x)$ and $c_y(y) \sim c_y(y) + 2\pi m_y(y)$,
where $m_x$ and $m_y$ are integer-valued piecewise constant functions.
Therefore, this symmetry should be regarded as a subsystem $\mathrm{U}(1)$ momentum symmetry.

The associated conservation law is nothing but the equation of motion~(\ref{cont_eom}), which reads
\begin{align}
    \label{momentum_conservation_law}
    \partial_{\tau}J^m_{\tau} = \partial_x\partial_y J^m_{xy},
\end{align}
where
\begin{align}
    \label{cont_momentum_current}
    J^{m}_{\tau} = i\mu_0\partial_{\tau}\phi,\;\; &J^{m}_{xy} = \frac{i}{\mu}\partial_{x}\partial_{y}\phi.
\end{align}
The conservation law~(\ref{momentum_conservation_law}) implies that one can define a momentum charge for a closed line $C_{x_0}$ lying on the plane $x=x_0$ as
\begin{align}
    \tilde{Q}_x^m(x_0;C_{x_0}) \coloneq \oint_{C_{x_0}} \left(\partial_{x}J^{m}_{xy} \, d\tau + J^{m}_{\tau} \, dy\right)
    = \oint_{C_{x_0}} \left(\frac{i}{\mu}\partial^2_{x}\partial_{y}\phi \, d\tau + i\mu_0\partial_{\tau}\phi \, dy\right).
\end{align}
This is formally quantized as $\tilde{Q}^{m}_{x}(x_0;C_{x_0}) \in \delta(0)\,\mathbb{Z}$, whose precise meaning should be understood in terms of the original UV lattice model.
If we instead consider the integrated charge $\int_{x_0}^{x_1} dx \, \tilde{Q}_x^m(x;C_{x})$ evaluated on a strip
$\{(\tau,x,y) \mid x\in[x_0,x_1],\, (\tau,x_0,y)\in C_{x_0}\}$,
it is quantized to integer values.
Similarly, one can define $\tilde{Q}^{m}_{y}(y_0;C_{y_0})$ for a closed loop $C_{y_0}$ lying on the plane $y=y_0$.

As mentioned above, one can deform the theory without breaking the momentum subsystem symmetry.
Under such deformations, the classical equation of motion is modified, which in turn leads to a modification of the momentum current.
Although we work with the undeformed kinetic term~(\ref{cont_action}) throughout this paper,
the discussions below, in particular those on Witten effects, can be applied to the deformed theories with appropriately redefined momentum charges.

\paragraph{Winding symmetry}
As in the case of the standard compact boson, the $\phi$-theory has a winding symmetry associated with the "smoothness" of field configurations.
Concretely, this symmetry originates from the identity
\begin{align}
    \label{cont_flatness}
    \partial_{\tau}(\partial_{x}\partial_{y}\phi) = \partial_{x}\partial_{y}(\partial_{\tau}\phi),
\end{align}
which can be rewritten as
\begin{align}
    \label{cont_winding_conservation}
    \partial_{\tau}J_{xy}^{w}=\partial_{x}\partial_{y}J^{w}_{\tau},
\end{align}
in terms of the gauge-invariant current operators
\begin{align}
    J^{w}_{\tau} = \frac{1}{2\pi}\partial_{\tau}\phi,\;\; J^{w}_{xy} = \frac{1}{2\pi} \partial_{x}\partial_{y}\phi.
\end{align}
This conservation law implies that one can define a winding number for a closed loop $C_{x_0}$ lying on the plane $x=x_0$ as
\begin{align}
    \tilde{Q}^{w}_{x}(x_0;C_{x_0}) \coloneq \oint_{C_{x_0}} \left(\partial_{x}J^{w}_{\tau} \, d \tau + J^{w}_{xy} \, dy\right)
    = \frac{1}{2\pi} \oint_{C_{x_0}} \left(\partial_{x}\partial_{\tau}\phi\, d\tau + \partial_{x}\partial_{y}\phi \, dy\right).
\end{align}
Again, this is formally quantized as $\tilde{Q}^{w}_{x}(x_0;C_{x_0}) \in \delta(0)\,\mathbb{Z}$, whose precise meaning should be understood in terms of the original UV lattice model.
If we instead consider the integrated charge $\int_{x_0}^{x_1} dx \, \tilde{Q}_x^w(x;C_{x})$ evaluated on a strip
$\{(\tau,x,y) \mid x\in[x_0,x_1],\, (\tau,x_0,y)\in C_{x_0}\}$,
it is quantized to integer values.
Similarly, one can define $\tilde{Q}^{w}_{y}(y_0;C_{y_0})$ for a closed loop $C_{y_0}$ lying on the plane $y=y_0$.

A charged operator for this symmetry is a vortex operator $e^{i\phi^{xy}}$, which creates a small spherical hole in the 3d spacetime with a boundary condition carrying a nontrivial winding number.
For example, inserting a vortex at $(\tau_0,x_0,y_0)$ leads to $\tilde{Q}^{w}_{x}(x;C_{x}) = \delta(x-x_0)$ for a closed loop 
$C_x = \{(\tau_0+ r_{\tau}\sin\theta,\,x,\,y_0 + r_{y}\cos\theta) \mid \theta \in [0,2\pi)\}$ with positive $r_{\tau}, r_{y}$.

\subsection{Modified Villain lattice model}

The modified Villain formulation enables us to control topological aspects of quantum field theories, such as monopoles and instantons.
This method has been widely used in the literature \cite{Sulejmanpasic:2019ytl,Gorantla:2021svj,Choi:2021kmx,Anosova:2022cjm,Anosova:2022yqx,Cheng:2022sgb,Fazza:2022fss,Jacobson:2023cmr,Berkowitz:2023pnz,Jacobson:2024hov,Honda:2024sdz,Honda:2024xmk,Chen:2024ddr,Jacobson:2024muj,Katayama:2025pmz,Seifnashri:2026ema,Aoki:2026pvq}.
For the $\phi$-theory, it reproduces an exact winding subsystem symmetry and its charged objects on the lattice,
which are absent in the original XY-plaquette model before taking the continuum limit.
Here, we briefly review the construction of \cite{Gorantla:2021svj}.

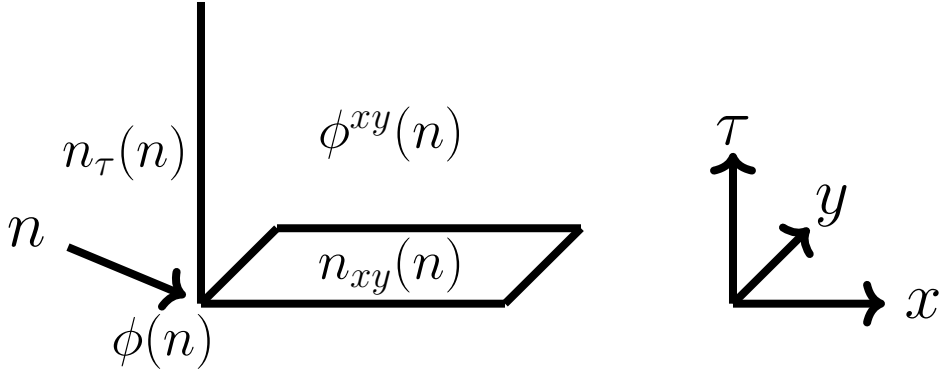
\begin{figure}[t]
    \centering
    \begin{tikzpicture}[scale=2.0]
	\begin{pgfonlayer}{nodelayer}
		\node [style=none] (0) at (-1, 1) {};
		\node [style=none] (1) at (-2, 0) {};
		\node [style=none] (2) at (2, 0) {};
		\node [style=none] (3) at (3, 1) {};
		\node [style=none] (4) at (-2, 4) {};
		\node [style=none] (5) at (-2, 0) {};
		\node [style=none] (6) at (0.5, 0.5) {\huge{$n_{xy}(n)$}};
		\node [style=none] (9) at (-3, 2) {\huge{$n_{\tau}(n)$}};
		\node [style=none] (10) at (-4.3, 0.95) {\Huge{$n$}};
		\node [style=none] (11) at (5, 0) {};
		\node [style=none] (12) at (6, 1) {};
		\node [style=none] (13) at (7, 0) {};
		\node [style=none] (14) at (5, 2) {};
		\node [style=none] (15) at (7.5, 0) {\Huge{$x$}};
		\node [style=none] (16) at (6.3, 1.3) {\Huge{$y$}};
		\node [style=none] (17) at (5, 2.3) {\Huge{$\tau$}};
		\node [style=none] (18) at (0.5, 2.25) {\huge{$\phi^{xy}(n)$}};
		\node [style=none] (19) at (-2.5, -0.5) {\huge{$\phi(n)$}};
		\node [style=none] (20) at (-3.75, 0.75) {};
		\node [style=none] (21) at (-2.2, 0.1) {};
	\end{pgfonlayer}
	\begin{pgfonlayer}{edgelayer}
		\draw [line width=3pt] (4.center) to (5.center);
		\draw [line width=3pt] (5.center) to (2.center);
		\draw [line width=3pt] (2.center) to (3.center);
		\draw [line width=3pt] (0.center) to (3.center);
		\draw [line width=3pt] (0.center) to (5.center);
		\draw [->, line width=3pt] (11.center) to (12.center);
		\draw [->, line width=3pt] (11.center) to (13.center);
		\draw [->, line width=3pt] (11.center) to (14.center);
		\draw [->, line width=3pt] (20.center) to (21.center);
	\end{pgfonlayer}
\end{tikzpicture}
    \caption{Variables in the modified Villain lattice model}
    \label{fig:villain-setup}
\end{figure}

We consider a three-dimensional cubic lattice with periodic boundary conditions.
The dynamical variables are as follows:
\begin{itemize}
    \item $\phi$ : real-valued variable living on sites,
    \item $\phi^{xy}$ : real-valued variable living on cubes,
    \item $n_{\tau}$ : integer-valued variable living on $\tau$-links,
    \item $n_{xy}$ : integer-valued variable living on $xy$-plaquettes.
\end{itemize}
For convenience, we label all variables by a site $n$, as shown in Fig.~\ref{fig:villain-setup}.

The lattice action is given by
\begin{align}
    S_0[\phi,n_{\tau},n_{xy}] = \frac{\beta_0}{2} \sum_{\text{$\tau$-link}}(\Delta_{\tau}\phi + 2\pi n_{\tau})^2 + \frac{\beta}{2}\sum_{\text{$xy$-plaquette}}(\Delta_x \Delta_y \phi + 2\pi n_{xy})^2 + i\sum_{\text{cube}} \phi^{xy}(\Delta_{\tau} n_{xy} - \Delta_x\Delta_y n_{\tau}).
\end{align}

We impose invariance under the following gauge transformations:
\begin{align}
    \label{2+1d_phi_theory_gauge_trsf}
    \begin{split}
    \phi &\to \phi + 2\pi k,\quad n_{\tau} \to n_{\tau} - \Delta_{\tau} k,\quad n_{xy} \to n_{xy} - \Delta_x\Delta_y k,\\
    \phi^{xy} &\to \phi^{xy} + 2\pi k',
    \end{split}
\end{align}
where $k$ and $k'$ are integer-valued gauge parameters defined on sites and cubes, respectively.

By fixing a gauge, we define the path integral measure as
\begin{align}
    \int\mathcal{D}\phi = \left(\prod_{n} \int_{-\pi}^{\pi} d\phi(n)\right) \left(\prod_{n} \sum_{n_{\tau}(n)\in\mathbb{Z}}\right) \left(\prod_{n}\sum_{n_{xy}(n)\in\mathbb{Z}}\right),
\end{align}
although the following analysis does not rely on gauge fixing.

\paragraph{Momentum symmetry}

The action $S_0$ possesses a momentum subsystem symmetry,
\begin{align}
    \label{2+1d_phi_theory_momentum_trsf}
    \phi(\tau,x,y) \to \phi(\tau,x,y) + c_x(x) + c_y(y),
\end{align}
where $c_x$ and $c_y$ are real-valued functions depending only on the $x$- and $y$-coordinates, respectively.
Due to the gauge redundancies~(\ref{2+1d_phi_theory_gauge_trsf}),
the transformation parameters are identified as
$c_x(x) \sim c_x(x) + 2\pi m_x(x)$ and $c_y(y) \sim c_y(y) + 2\pi m_y(y)$,
where $m_x$ and $m_y$ are integer-valued functions.
Therefore, this symmetry is a $\mathrm{U}(1)$ subsystem symmetry.

The corresponding conservation law follows from the equation of motion for $\phi$:
\begin{align}
    \label{lat_momentum_conservation}
    \Dtau J^{m}_{\tau}(n-\tauhat) = \Dx\Dy J^{m}_{xy}(n-\xhat-\yhat),
\end{align}
where the current operators are given by
\begin{align}
    \label{lat_momentum_current}
    J^{m}_{\tau} &= i\beta_0 (\Dtau \phi + 2\pi n_{\tau}),\\
    J^{m}_{xy} &= i\beta (\Dx\Dy\phi + 2\pi n_{xy}).
\end{align}

This conservation law allows us to define symmetry charge operators. For example, we define
\begin{align}
    \tilde{Q}^{m}_x(x_0;C_{x_0}) = \sum_{\text{$\tau$-link} \in C_{x_0}} J^{m}_{\tau} 
    + \sum_{\text{$y$-link} \in C_{x_0}} \Dx J^{m}_{xy},
\end{align}
where $C_{x_0}$ is a closed loop on the dual lattice within the plane $x = x_0$.
Similarly, one can define a charge $\tilde{Q}^{m}_y(y_0;C_{y_0})$ associated with a closed loop $C_{y_0}$ on the plane $y = y_0$.

\paragraph{Winding symmetry}

The theory is invariant under the transformation
\begin{align}
    \phi^{xy}(\tau,x,y) \to \phi^{xy}(\tau,x,y) + \tilde{c}_x(x) + \tilde{c}_y(y),
\end{align}
where $\tilde{c}_x$ and $\tilde{c}_y$ are real-valued functions depending only on the $x$- and $y$-coordinates, respectively.
Again, these parameters have a $2\pi$ periodicity due to the gauge redundancy~(\ref{2+1d_phi_theory_gauge_trsf}).
This defines a winding $\mathrm{U}(1)$ subsystem symmetry.

The Lagrange multiplier $\phi^{xy}$ imposes the flatness condition
\begin{align}
    \label{2+1d_phi_theory_flatness}
    \Dtau \nxy - \Dx\Dy \ntau = 0.
\end{align}

An example of a charged operator for this symmetry is $e^{i\phi^{xy}}(\tau,x,y)$, which carries unit charge.
If we insert the operator $e^{i\phi^{xy}(n_0)}$ in the path integral and integrate out $\phi^{xy}$,
the flatness condition~(\ref{2+1d_phi_theory_flatness}) is modified as
\begin{align}
    \Dtau \nxy(n_0) - \Dx\Dy \ntau(n_0) = +1.
\end{align}
Thus, $e^{i\phi^{xy}}$ creates a vortex with nontrivial winding, and we refer to it as a vortex operator.

The conservation law for this winding symmetry is equivalent to the flatness condition~(\ref{2+1d_phi_theory_flatness}).
Using the gauge-invariant current operators
\begin{align}
    J^{w}_{\tau} &= \frac{1}{2\pi}(\Dtau \phi + 2\pi \ntau),\\
    J^{w}_{xy} &= \frac{1}{2\pi}(\Dx\Dy\phi + 2\pi \nxy),
\end{align}
we can rewrite~(\ref{2+1d_phi_theory_flatness}) as
\begin{align}
    \Dtau J^{w}_{xy}(n) = \Dx\Dy J^{w}_{\tau}(n),
\end{align}
which is the lattice analogue of~(\ref{cont_winding_conservation}).

Accordingly, one can define a winding charge evaluated on a closed loop $C_{x_0}$ on the dual lattice of the plane $x = x_0+\frac{1}{2}$:
\begin{align}
    \tilde{Q}^{w}_x(x_0;C_{x_0}) = \sum_{\text{$\tau x$-plaquette} \in C_{x_0}} \Dx J^{w}_{\tau} 
    + \sum_{\text{$xy$-plaquette} \in C_{x_0}} J^{w}_{xy}.
\end{align}
Similarly, one can define a winding charge $\tilde{Q}^{w}_y(y_0;C_{y_0})$ associated with a closed loop $C_{y_0}$ on the plane $y = y_0+\frac{1}{2}$.

\section{Lattice construction}
\label{section_lattice}
In this section, we begin our study of theta terms in the 2+1d $\phi$-theory by constructing their lattice realizations based on the modified Villain formulation.
The corresponding continuum descriptions will be discussed in the next section.

\subsection{Bulk theta term}
\label{sec_lat_bulk}

To construct a theta term, we need a topological charge built from the integer-valued fields in the modified Villain lattice model.
Our task is therefore to identify a gauge-invariant combination of these integer-valued fields under the flatness condition~(\ref{2+1d_phi_theory_flatness}).

One such gauge-invariant combination is
\begin{align}
    \label{lat_topo_charge}
    Q_{\mathrm{bulk}}[\ntau,\nxy] = \sum_{n} \bigl(\ntau(n)\nxy(n+\tauhat) + \nxy(n)\ntau(n+\xhat+\yhat)\bigr).
\end{align}
We note that this can be regarded as a fractonic analogue of the cup product in algebraic topology.\footnote{This is essentially equivalent to the construction presented in the appendix of Ref.~\cite{Cao:2023doz}, where $\mathbb{Z}_2$ subsystem symmetry is gauged using the dual lattice.}
\footnote{Consider 1-cochains $\alpha,\beta\in C^{1}(T^2;\mathbb{Z})$ defined on a 2d square lattice. Their cup product is given by $(\alpha \cup \beta)_{xy}(n) = \alpha_x(n)\beta_y(n+\xhat) - \alpha_y(n)\beta_x(n+\yhat)$.}
For comparison, in Appendix~\ref{Appendix_topological_charge}, we present an alternative construction of the topological charge for admissible field configurations in the original XY-plaquette model, which does not rely on the modified Villain formulation.

Let us consider a concrete field configuration
\begin{align}
    \begin{split}
    \label{lattice_sample_config}
    \ntau(\tau,x,y) &= \left\{
        \begin{array}{ll}
            \delta_{\tau,\tau_0} & x_0 \leq x \leq x_0 + r_x,\\
            0 & \text{otherwise},
        \end{array}
    \right.\\
    \nxy(\tau,x,y) &= \delta_{x,x_1}\delta_{y,y_1},
    \end{split}
\end{align}
which satisfies the flatness condition~(\ref{2+1d_phi_theory_flatness}) for any $r_x > 0$ and integers $x_0,y_0,x_1,y_1$.

We then obtain
\begin{align}
    \begin{split}
        \label{lattice_sample_config_topo}
        \frac{1}{2\pi}\sum_{\tau} (\DtauPhi)(\tau,X,Y) &=
        \begin{cases}
            1 & x_0 \leq X \leq x_0 + r_x,\\
            0 & \text{otherwise},
        \end{cases}\\
        \frac{1}{2\pi}\sum_{X_0\leq x\leq X_1} \sum_{y} (\DxyPhi)(T,x,y) &=
        \begin{cases}
            1 & X_0 \leq x_1 \leq X_1,\\
            0 & \text{otherwise},
        \end{cases}\\
        \frac{1}{2\pi}\sum_{Y_0\leq y\leq Y_1} \sum_{x} (\DxyPhi)(T,x,y) &=
        \begin{cases}
            1 & Y_0 \leq y_1 \leq Y_1,\\
            0 & \text{otherwise}.
        \end{cases}
    \end{split}
\end{align}

The topological charge for this configuration is given by
\begin{align}
    \label{lat_sample_topo_charge}
    Q_{\mathrm{bulk}}[\ntau,\nxy] =
        \begin{cases}
            2 & x_0 \leq x_1 \leq x_0 + r_x - 1,\\
            1 & x_1 = x_0 - 1,\; x_0 + r_x,\\
            0 & \text{otherwise}.
        \end{cases}
\end{align}

As this example illustrates, $Q_{\mathrm{bulk}}[\ntau,\nxy]$ typically takes even integer values, while certain configurations yield odd values.
We will discuss its continuum counterpart in the next section.

We now define the bulk theta term\footnote{This expression is obtained by the replacements $2\pi n_{\tau}\to \Dtau\phi + 2\pi n_{\tau}$ and $2\pi n_{xy} \to \Dx\Dy \phi + 2\pi n_{xy}$ in Eq.~(\ref{lat_topo_charge}).}
\begin{align}
    \begin{split}
    \label{lat_bulk_theta}
    \sbulk[\btheta;\phi,\ntau,\nxy] = -\frac{i\btheta}{(2\pi)^2} &\sum_{n}\left\{(\DtauPhi)(n)(\DxyPhi)(n+\tauhat)\right. \\
    &\left.+ (\DxyPhi)(n) (\DtauPhi)(n+\xhat+\yhat)\right\},
    \end{split}
\end{align}
which is manifestly gauge invariant.
It can be rewritten as
\begin{align}
    \label{lat_rewrite_bulk_theta}
    \begin{split}
    \sbulk[\btheta;\phi,\ntau,\nxy] =& -i\btheta Q_{\mathrm{bulk}}[\ntau,\nxy] \\
    &+\frac{i\btheta}{2\pi} \sum_{n} (\Dtau \nxy - \Dx\Dy \ntau)(n) \left(\phi(n+\tauhat+\xhat+\yhat) + \phi(n) \right).
    \end{split}
\end{align}
Under the flatness condition~(\ref{2+1d_phi_theory_flatness}), this term depends only on the topological charge~(\ref{lat_topo_charge}).
Therefore, the theta term $\sbulk$ is topological in the sense that it does not affect the classical equations of motion.
One also finds the periodicity of the theta angle, namely $\btheta \sim \btheta + 2\pi$.

\begin{figure}[t]
    \centering
    \begin{tikzpicture}
	\begin{pgfonlayer}{nodelayer}
		\node [style=none] (0) at (-3, 2) {};
		\node [style=none] (1) at (-3, -2) {};
		\node [style=none] (2) at (1, -2) {};
		\node [style=none] (3) at (1, 2) {};
		\node [style=none] (4) at (-1, 3) {};
		\node [style=none] (5) at (3, 3) {};
		\node [style=none] (6) at (3, -1) {};
		\node [style=none] (7) at (-1, -1) {};
		\node [style=red diamond] (8) at (0, 0.5) {};
		\node [style=blue dot] (9) at (-3, -2) {};
		\node [style=blue dot] (10) at (3, 3) {};
		\node [style=none] (11) at (4.5, 3) {$-\frac{\btheta}{2\pi}$};
		\node [style=none] (12) at (-4.5, -2) {$-\frac{\btheta}{2\pi}$};
		\node [style=none] (13) at (5, -2) {};
		\node [style=none] (14) at (5, 0) {};
		\node [style=none] (15) at (7, -2) {};
		\node [style=none] (16) at (6, -1.5) {};
		\node [style=none] (17) at (5, 0.5) {$\tau$};
		\node [style=none] (18) at (6.5, -1.25) {$y$};
		\node [style=none] (19) at (7.5, -2) {$x$};
	\end{pgfonlayer}
	\begin{pgfonlayer}{edgelayer}
		\draw (0.center) to (1.center);
		\draw (1.center) to (2.center);
		\draw (2.center) to (6.center);
		\draw (3.center) to (2.center);
		\draw (3.center) to (5.center);
		\draw (5.center) to (6.center);
		\draw (4.center) to (5.center);
		\draw (4.center) to (0.center);
		\draw (0.center) to (3.center);
		\draw [style=dashedline] (1.center) to (7.center);
		\draw [style=dashedline] (7.center) to (6.center);
		\draw [style=dashedline] (4.center) to (7.center);
		\draw [style=arrowline] (15.center) to (13.center);
		\draw [style=arrowline] (14.center) to (13.center);
		\draw [style=arrowline] (16.center) to (13.center);
	\end{pgfonlayer}
\end{tikzpicture}
    \caption{In the presence of the bulk theta term, a vortex operator (red diamond) induces fractional $\phi$ excitations (blue circles).}
    \label{fig:bulk-theta}
\end{figure}
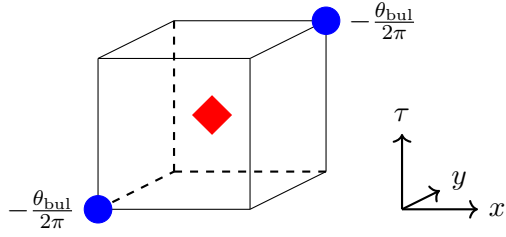

However, in the presence of a vortex at a site $n$, the additional term in the second line of~(\ref{lat_rewrite_bulk_theta}) contributes,
implying that the vortex is dressed by $\phi$ fields carrying momentum charges in the presence of the bulk theta term (Fig.~\ref{fig:bulk-theta}).
As we discuss below, this leads to a Witten effect.

\paragraph{Witten effect}

Let us derive the Witten effect~\cite{Witten:1979ey} induced by the bulk theta term using the Ward--Takahashi identity associated with the momentum subsystem symmetry.
Witten effects on the lattice have been studied in \cite{Sulejmanpasic:2019ytl,Anosova:2022cjm,Abe:2023uan,Aoki:2023lqp,Onoda:2025gqa,Katayama:2025pmz}.

We first consider a shift transformation $\phi \to \phi + \alpha$, where $\alpha$ is a real-valued function defined on lattice sites.
The kinetic term transforms as
\begin{align}
    \begin{split}
    \label{change_of_action}
    &S_0[\phi + \alpha,\ntau,\nxy] - S_0[\phi,\ntau,\nxy] \\
    =& \sum_{n} \left\{\beta_0 \Dtau \alpha(n) (\Dtau\phi+2\pi\ntau)(n) + \beta \Dx\Dy\alpha(n) (\Dx\Dy\phi + 2\pi\nxy)(n)\right\}\\
    & + \sum_{n} \left\{\frac{\beta_0}{2}(\Dtau \alpha(n))^2 + \frac{\beta}{2} (\Dx\Dy\alpha(n))^2\right\}\\
    =& \sum_{n} \alpha(n) \left\{-\beta_0 (\Dtau(\Dtau \phi + 2\pi \ntau))(n-\tauhat) + \beta \Dx\Dy(\Dx\Dy\phi+2\pi\nxy)(n-\xhat-\yhat)\right\} + O(\alpha^2)\\
    =& -i \sum_{n} \alpha(n) \left\{-\Dtau J^{m}_{\tau} (n-\tauhat) + \Dx\Dy J^{m}_{xy}(n-\xhat-\yhat)\right\} + O(\alpha^2),
    \end{split}
\end{align}
where we used the momentum current operators~(\ref{lat_momentum_current}).
In the second equality, we used a lattice analogue of integration by parts.
In the absence of the theta term, this implies the operator equation~(\ref{lat_momentum_conservation}) at each site $n$.

The momentum charge operator evaluated on an $xy$-plane with a modulation $f(x)$ is given by
\begin{align}
    Q_{x,f}^{m}(\tau) = \sum_{x,y} f(x) J^{m}_{\tau}(\tau,x,y).
\end{align}

Let us consider
\begin{align}
    \alpha_f(\tau,x,y) = \begin{cases}
        \alpha_0 f(x) & \tau_0 < \tau \leq \tau_1, \\
        0 & \text{otherwise}.
    \end{cases}
\end{align}
In this case, using Eq.~(\ref{change_of_action}), one finds
\begin{align}
    \label{def_of_momentum_charge_without_theta}
    Q_{x,f}^{m}(\tau_1) - Q_{x,f}^{m}(\tau_0) = \frac{1}{i}\left.\frac{d}{d\alpha_0} S_0[\phi+\alpha_f,\ntau,\nxy]\right|_{\alpha_0=0}.
\end{align}

On the other hand, from Eq.~(\ref{lat_rewrite_bulk_theta}), we obtain
\begin{align}
    \begin{split}
        \label{change_of_bulk_theta}
        &\left.\frac{d}{d\alpha_0}\sbulk[\btheta;\phi+\alpha_f,\ntau,\nxy]\right|_{\alpha_0=0}\\
        =& \frac{i\btheta}{2\pi} \sum_{\tau_0 < \tau \leq \tau_1}\sum_{x,y} (\Dtau \nxy - \Dx\Dy \ntau)(\tau,x,y) f(x) \\
        & + \frac{i\btheta}{2\pi} \sum_{\tau_0 \leq \tau < \tau_1}\sum_{x,y} (\Dtau \nxy - \Dx\Dy \ntau)(\tau,x,y) f(x+1).
    \end{split}
\end{align}

We now insert a vortex operator $e^{i\phi^{xy}}$ at the origin $(0,0,0)$.
If $\tau_0 < 0 < \tau_1$, one finds
\begin{align}
    \begin{split}
        \label{lat_bulk_Witten}
        & \Braket{Q_{x,f}^{m}(\tau_1)e^{i\phi^{xy}(0,0,0)}\cdots} - \Braket{e^{i\phi^{xy}(0,0,0)}Q_{x,f}^{m}(\tau_0)\cdots} \\
        =& \frac{i}{Z} \frac{d}{d\alpha_0}\left.\left[\int \mathcal{D}\phi\, e^{-S_0[\phi+\alpha_f,\ntau,\nxy]-\sbulk[\btheta;\phi,\ntau,\nxy]} e^{i\phi^{xy}(0,0,0)} \cdots\right]\right|_{\alpha_0=0}\\
        =& \frac{i}{Z} \frac{d}{d\alpha_0}\left.\left[\int \mathcal{D}\phi\, e^{-S_0[\phi,\ntau,\nxy]-\sbulk[\btheta;\phi-\alpha_f,\ntau,\nxy]} e^{i\phi^{xy}(0,0,0)} \cdots\right]\right|_{\alpha_0=0}\\
        =& - \frac{\btheta}{2\pi} \left(f(0)+f(1)\right)\Braket{e^{i\phi^{xy}(0,0,0)}\cdots},
    \end{split}
\end{align}
where $Z$ is the partition function of the theory,
and $\cdots$ denotes other operator insertions far from the region $\tau_0 < \tau < \tau_1$.
The last equality follows from $(\Dtau \nxy - \Dx\Dy \ntau)(0,0,0) = +1$, which arises after integrating out $\phi^{xy}$ in the presence of the vortex operator $e^{i\phi^{xy}(0,0,0)}$.
If $f(x)=1$, Eq.~(\ref{lat_bulk_Witten}) implies that a vortex induces a fractional momentum charge $-\frac{\btheta}{\pi}$.
This is the Witten effect.

\subsection{Foliated theta term}
\label{sec_lat_foliated}

In this section, we introduce the foliated theta term.
To illustrate the construction, we begin with two 1+1d compact bosons $\phi_1, \phi_2$,
with the identifications $\phi_1 \sim \phi_1 + 2\pi$ and $\phi_2 \sim \phi_2 + 2\pi$.
In this normalization, there is a topological charge $\frac{1}{(2\pi)^2}\int d\phi_1 \wedge d\phi_2 \in \mathbb{Z}$,
where $d\phi_1$ and $d\phi_2$ are the currents associated with the two winding symmetries.
This becomes trivial if $\phi_1 = \phi_2$.

In the $\phi$-theory, the winding subsystem symmetry gives rise to an independent winding current on each $\tau y$-plane.
This structure allows us to couple neighboring planes in an analogous manner.\footnote{One can similarly construct the coupling for $\tau x$-planes.}
Motivated by this, we define the foliated theta term
\begin{align}
    \begin{split}
    \label{lat_foliated_theta}
    &\sfol[\ftheta;\phi,\ntau,\nxy] \\
    =& +\frac{i}{(2\pi)^2}\sum_{x} \ftheta(x)\sum_{\tau,y} \{\Dx(\DtauPhi)(n) (\DxyPhi)(n-\hat{x}+\hat{\tau}) \\
    &- (\DxyPhi)(n) \Dx(\DtauPhi)(n-\hat{x}+\hat{y})\},
    \end{split}
\end{align}
where $n$ denotes a site specified by $(\tau,x,y)$.
We emphasize that the theta parameter $\ftheta(x)$ can depend on $x$.

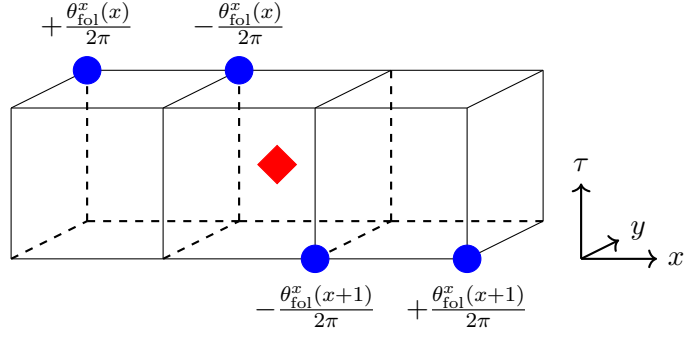
\begin{figure}[t]
    \centering
    \begin{tikzpicture}
	\begin{pgfonlayer}{nodelayer}
		\node [style=none] (0) at (-3, 2) {};
		\node [style=none] (1) at (-3, -2) {};
		\node [style=none] (2) at (1, -2) {};
		\node [style=none] (3) at (1, 2) {};
		\node [style=none] (4) at (-1, 3) {};
		\node [style=none] (5) at (3, 3) {};
		\node [style=none] (6) at (3, -1) {};
		\node [style=none] (7) at (-1, -1) {};
		\node [style=none] (8) at (-7, 2) {};
		\node [style=none] (9) at (-7, -2) {};
		\node [style=none] (10) at (-5, 3) {};
		\node [style=none] (11) at (5, -2) {};
		\node [style=none] (12) at (7, -1) {};
		\node [style=none] (13) at (7, 3) {};
		\node [style=none] (14) at (5, 2) {};
		\node [style=none] (15) at (-5, -1) {};
		\node [style=red diamond] (16) at (0, 0.5) {};
		\node [style=blue dot] (17) at (-5, 3) {};
		\node [style=blue dot] (18) at (-1, 3) {};
		\node [style=blue dot] (19) at (1, -2) {};
		\node [style=blue dot] (20) at (5, -2) {};
		\node [style=none] (21) at (1, -3.25) {$-\frac{\ftheta(x+1)}{2\pi}$};
		\node [style=none] (22) at (5, -3.25) {$+\frac{\ftheta(x+1)}{2\pi}$};
		\node [style=none] (23) at (-1, 4.25) {$-\frac{\ftheta(x)}{2\pi}$};
		\node [style=none] (24) at (-5, 4.25) {$+\frac{\ftheta(x)}{2\pi}$};
		\node [style=none] (25) at (8, -2) {};
		\node [style=none] (26) at (8, 0) {};
		\node [style=none] (27) at (10, -2) {};
		\node [style=none] (28) at (9, -1.5) {};
		\node [style=none] (29) at (8, 0.5) {$\tau$};
		\node [style=none] (30) at (9.5, -1.25) {$y$};
		\node [style=none] (31) at (10.5, -2) {$x$};
	\end{pgfonlayer}
	\begin{pgfonlayer}{edgelayer}
		\draw (0.center) to (1.center);
		\draw (1.center) to (2.center);
		\draw [style=dashedline] (2.center) to (6.center);
		\draw (3.center) to (2.center);
		\draw (3.center) to (5.center);
		\draw [style=dashedline] (5.center) to (6.center);
		\draw (4.center) to (5.center);
		\draw (4.center) to (0.center);
		\draw (0.center) to (3.center);
		\draw [style=dashedline] (1.center) to (7.center);
		\draw [style=dashedline] (7.center) to (6.center);
		\draw [style=dashedline] (4.center) to (7.center);
		\draw (3.center) to (14.center);
		\draw (14.center) to (11.center);
		\draw (2.center) to (11.center);
		\draw (11.center) to (12.center);
		\draw (12.center) to (13.center);
		\draw (13.center) to (14.center);
		\draw (5.center) to (13.center);
		\draw [style=dashedline] (6.center) to (12.center);
		\draw [style=dashedline] (10.center) to (15.center);
		\draw [style=dashedline] (15.center) to (7.center);
		\draw [style=dashedline] (9.center) to (15.center);
		\draw (9.center) to (1.center);
		\draw (8.center) to (9.center);
		\draw (8.center) to (0.center);
		\draw (8.center) to (10.center);
		\draw (10.center) to (4.center);
		\draw [style=arrowline] (27.center) to (25.center);
		\draw [style=arrowline] (26.center) to (25.center);
		\draw [style=arrowline] (28.center) to (25.center);
	\end{pgfonlayer}
\end{tikzpicture}
    \caption{In the presence of the foliated theta term, a vortex operator (red diamond) induces fractional $\phi$ excitations (blue circles).}
    \label{fig:foliated-theta}
\end{figure}

Eq.~(\ref{lat_foliated_theta}) is manifestly gauge invariant, and it can be rewritten as
\begin{align}
    \begin{split}
    &\sfol[\ftheta;\phi,\ntau,\nxy] \\
    =& +i \sum_{x} \ftheta(x)\sum_{\tau,y} \{\Dx\ntau(n)\, \nxy(n-\hat{x}+\hat{\tau}) - \nxy(n)\, \Dx\ntau(n-\hat{x}+\hat{y})\} \\
    &- i \sum_{\tau,x,y}(\Dtau \nxy - \Dx\Dy\ntau)(n)\left\{\frac{\ftheta(x+1)}{2\pi}\Dx\phi(n+\xhat)-\frac{\ftheta(x)}{2\pi}\Dx\phi(n+\hat{\tau}-\hat{x}+\hat{y})\right\}.
    \end{split}
\end{align}
When there are no vortices, and hence the flatness condition~(\ref{2+1d_phi_theory_flatness}) is satisfied, this term depends only on the topological sector of the configuration.
However, if vortices are introduced, the last line contributes,
implying that the vortex is dressed by $\phi$ fields carrying momentum charges in the presence of the foliated theta term (Fig.~\ref{fig:foliated-theta}).
As we discuss below, this leads to a generalized Witten effect.

\paragraph{Witten effect}

One can derive the Witten effect for the foliated theta term in the same manner as in the previous section.
Using
\begin{align}
    \begin{split}
        \label{change_of_foliated_theta}
        &\left.\frac{d}{d\alpha_0}\sfol[\ftheta;\phi+\alpha_f,\ntau,\nxy]\right|_{\alpha_0=0}\\
        =& - i \sum_{\tau_0 <\tau\leq\tau_1}\sum_{x,y}(\Dtau \nxy - \Dx\Dy\ntau)(\tau,x,y) \frac{\ftheta(x+1)}{2\pi}\Dx f(x+1) \\
        & + i \sum_{\tau_0 \leq\tau < \tau_1}\sum_{x,y}(\Dtau \nxy - \Dx\Dy\ntau)(\tau,x,y) \frac{\ftheta(x)}{2\pi}\Dx f(x-1) ,
    \end{split}
\end{align}
we obtain
\begin{align}
        \label{lat_foliated_Witten}
        \begin{split}
        &\Braket{Q_{x,f}^{m}(\tau_1)e^{i\phi^{xy}(0,0,0)}\cdots} - \Braket{e^{i\phi^{xy}(0,0,0)}Q_{x,f}^{m}(\tau_0)\cdots}\\
        = & \left(\frac{\ftheta(1)}{2\pi}\Dx f(1)-\frac{\ftheta(0)}{2\pi}\Dx f(-1)\right) \Braket{e^{i\phi^{xy}(0,0,0)}\cdots}.
        \end{split}
\end{align}

To compare with the continuum description in Sec.~\ref{sec_cont_foliated}, we embed the lattice into continuous spacetime.
Let us compute the leading behavior of Eq.~(\ref{lat_foliated_Witten}) in the lattice spacing $a_x$ along the $x$-direction.
We shift the $x$-coordinates of $\ftheta$ and $f$ so that the vortex in Fig.~\ref{fig:foliated-theta} is located on the plane $x=0$.
Then we obtain the momentum charge of the vortex operator
\begin{align}
    \begin{split}
        \label{lat_foliated_Witten_shifted}
    Q_{x,f}^{m,\mathrm{shifted}} &= \frac{1}{2\pi}\ftheta\left(\frac{a_x}{2}\right) \left\{f\left(\frac{3}{2}a_x\right) - f\left(\frac{a_x}{2}\right)\right\} - \frac{1}{2\pi}\ftheta\left(-\frac{a_x}{2}\right) \left\{f\left(-\frac{3}{2}a_x\right) - f\left(-\frac{a_x}{2}\right)\right\}\\
    &= \frac{a_x^2}{2\pi} \left(\ftheta{}^{'}(0) f'(0) + 2\ftheta(0)f''(0)\right) + O(a_x^4).
    \end{split}
\end{align}
If $\ftheta(x)$ is a nonzero constant, the vortex does not induce a net momentum charge or dipole moment,
but instead induces a quadrupole moment of order $O(a_x^2)$.
If $\ftheta(x) \sim 1$ varies smoothly, the vortex induces a dipole moment along the $x$-direction, again of order $O(a_x^2)$.
Although these effects vanish in the strict continuum limit $a_x \to 0$,
they are robust lattice effects under any deformations preserving the momentum and winding subsystem symmetries.

\section{Continuum description}
\label{section_continuum}

We now move on to the continuum description of the theta terms constructed on the lattice in the previous section.

\subsection{Bulk theta term}
\label{sec_cont_bulk}

We study the continuum description of the bulk theta term in the $\phi$-theory,
which was first introduced in \cite{Bedogna:2026bck}.
The bulk theta term is given by
\begin{align}
    \label{cont_bulk_theta}
    \sbulk[\btheta;\phi] = -\frac{i\btheta}{2\pi^2} \int d\tau\, dx\, dy\, \partial_{\tau} \phi\, \partial_{x}\partial_{y}\phi.
\end{align}
The associated topological charge is
\begin{align}
    \label{cont_topological_charge}
    Q_{\mathrm{bulk}} = \frac{1}{2\pi^2} \int d\tau\, dx\, dy\, \partial_{\tau} \phi\, \partial_{x}\partial_{y}\phi.
\end{align}

For the standard compact boson, in which $\partial_{\tau}\phi$, $\partial_{x}\phi$, and $\partial_{y}\phi$ are all well-defined, 
the topological term reduces to a total derivative and hence becomes trivial:
\begin{align}
    Q_{\mathrm{bulk}} = \frac{1}{(2\pi)^2} \int d\tau\, dx\, dy\, \left\{\partial_{x} (\partial_{\tau}\phi\,\partial_{y}\phi) - \partial_{\tau} (\partial_{x}\phi\,\partial_{y}\phi) + \partial_{y} (\partial_{\tau}\phi\,\partial_{x}\phi)\right\} = 0.
\end{align}
In contrast, in the $\phi$-theory, which admits discontinuous field configurations, 
$\partial_{x}\phi$ and $\partial_{y}\phi$ are no longer well-defined, and the theta term can become nontrivial.

In Appendix~\ref{appendix_quantization_of_topological_charge}, we show that $Q_{\mathrm{bulk}} \in \mathbb{Z}$ on a three-dimensional torus,
which implies the periodicity of $\btheta$, namely $\btheta \sim \btheta + 2\pi$.
\footnote{The normalization factor in Eq.~(\ref{cont_bulk_theta}) differs from that in \cite{Bedogna:2026bck} by a factor of $2$. This difference is related to a subtle issue in the quantization of the topological charge, which we discuss below.}

We consider the field configuration
\begin{align}
    \label{cont_sample_config}
    \phi(\tau,x,y) = 2\pi\left(\frac{x}{\ell_{x}}\Theta(y-y_1) + \frac{y}{\ell_{y}}\Theta(x-x_1)-\frac{xy}{\ell_x \ell_y}\right)
    + 2\pi \frac{\tau}{\ell_\tau} \Theta(x-x_0) \Theta((x_0+r_x)-x),
\end{align}
where $x_0,x_1,y_1,r_x$ are real numbers, and in particular $r_x>0$.
The parameters $\ell_{\mu}$ denote the system size in the $\mu$-direction, namely $\tau \sim \tau + \ell_{\tau}$, $x \sim x + \ell_{x}$, and $y \sim y + \ell_{y}$.
For each $\mu=\tau,x,y$, the field $\phi$ on the two planes $x_\mu = 0$ and $x_\mu = \ell_{\mu}$ is glued by appropriate transition functions.

One can verify that
\begin{align}
    \begin{split}
        \label{cont_sample_config_topo}
        \frac{1}{2\pi} \oint d \tau \, \partial_{\tau} \phi(\tau,X,Y) &=
        \begin{cases}
            1 & x_0 \leq X \leq x_0 + r_x,\\
            0 & \text{otherwise},
        \end{cases}\\
        \frac{1}{2\pi}\int_{X_0}^{X_1} dx \oint dy\, \partial_x \partial_y \phi(T,x,y) &=
        \begin{cases}
            1 & X_0 \leq x_1 \leq X_1,\\
            0 & \text{otherwise},
        \end{cases}\\
        \frac{1}{2\pi}\int_{Y_0}^{Y_1} dy \oint dx\, \partial_x \partial_y \phi(T,x,y) &=
        \begin{cases}
            1 & Y_0 \leq y_1 \leq Y_1,\\
            0 & \text{otherwise}.
        \end{cases}
    \end{split}
\end{align}

Comparing with Eq.~(\ref{lattice_sample_config_topo}), this configuration is the continuum analogue of Eq.~(\ref{lattice_sample_config}).
For this configuration, we obtain the formal expression for the topological charge
\begin{align}
    Q_{\mathrm{bulk}} = 2 \int dx \, \Theta(x-x_0) \Theta((x_0+r_x)-x) \delta(x-x_1).
\end{align}

To ensure that this quantity is quantized to integer values, we must treat carefully the cases $x_1 = x_0$ and $x_1 = x_0 + r_x$, which require a regularization.\footnote{In Eq.~(\ref{cont_sample_config_topo}), we have been somewhat schematic about this point.}
The issue is how to treat the discontinuities of the step functions arising from the transition functions defining $\phi$.
Here, we adopt the prescription
\begin{align}
    \label{our_prescription}
    \int dx\, \Theta(x-x_0) \delta(x-x_0) = \frac{1}{2}.
\end{align}

To justify this, we return to the modified Villain lattice model.
Recall that Eq.~(\ref{lattice_sample_config}) provides the lattice analogue of the configuration~(\ref{cont_sample_config}).
With the prescription~(\ref{our_prescription}), we obtain
\begin{align}
    Q_{\mathrm{bulk}} =
    \begin{cases}
        2 & x_0 < x_1 < x_0 + r_x,\\
        1 & x_1 = x_0,\; x_0 + r_x,\\
        0 & \text{otherwise},
    \end{cases}
\end{align}
which precisely matches the lattice result~(\ref{lat_sample_topo_charge}).
An alternative justification of Eq.~(\ref{our_prescription}) is to require that the Leibniz rule holds for the squared step function:
\begin{align}
    \label{Leibniz}
    \partial_{x} \left(\Theta(x-x_0)\right)^{2} = 2 \Theta(x-x_0) \delta(x-x_0).
\end{align}
Integrating both sides, we obtain
\begin{align}
    1 = 2 \int dx\, \Theta(x-x_0) \delta(x-x_0),
\end{align}
which implies Eq.~(\ref{our_prescription}).

This example implies that the topological charge typically takes even integer values, while only special configurations with accidental coincidences (such as $x_1 = x_0$ or $x_1 = x_0 + r_x$ in this case) yield odd values.
In this sense, the periodicity $\btheta \sim \btheta + \pi$ is broken in a subtle manner.

This can also be understood from Fig.~\ref{fig:bulk-theta}.
The momentum charge induced by the Witten effect is split into two contributions.
While this splitting is not observable macroscopically, at the lattice level the subsystem momentum charges remain fractional when $\btheta = \pi$.
Thus, the periodicity $\btheta \sim \btheta + \pi$ is violated in a subtle way on the lattice.
In particular, there appears to be no lattice realization with the periodicity $\btheta \sim \btheta + \pi$.

Under the shift transformation $\phi \to \phi + \alpha$, the bulk theta term changes as
\begin{align}
    \label{change_of_bulk_theta}
    \sbulk[\btheta; \phi + \alpha] - \sbulk[\btheta; \phi] = \frac{i \btheta}{2\pi^2} \int d\tau\, dx\, dy\, \alpha \left\{\partial_{\tau}(\partial_{x}\partial_{y}\phi) - \partial_{x}\partial_{y}(\partial_{\tau}\phi)\right\}.
\end{align}
In the absence of winding vortices, this vanishes due to Eq.~(\ref{cont_flatness}).
In this sense, $\sbulk$ is a topological term and does not affect the classical equations of motion.

\paragraph{Witten effect}

Let us derive the Witten effect.
We first consider a shift transformation $\phi \to \phi + \alpha$,
where $\alpha$ is a real-valued function that may depend on spacetime points.
The kinetic term transforms as
\begin{align}
    \begin{split}
    \label{cont_change_of_action}
    &S_0[\phi+\alpha] - S_0[\phi] \\
    =& -i \int d\tau dx dy\, \left(\partial_{\tau} \alpha\, J_{\tau}^m + \partial_{x}\partial_{y}\alpha\, J_{xy}^m\right) + O(\alpha^2)\\
    =& i \int d\tau dx dy\, \alpha \left(\partial_{\tau} J_{\tau}^m - \partial_{x}\partial_{y} J_{xy}^m\right) + O(\alpha^2),
    \end{split}
\end{align}
which implies Eq.~(\ref{momentum_conservation_law}) in the absence of the theta term.

As mentioned in Section~\ref{review_of_continuum}, one can deform the theory while preserving the momentum subsystem symmetry.
However, the momentum symmetry requires that $\phi$ appears in the action only through the combinations $\partial_{\tau}\phi$ and $\partial_{x}\partial_{y}\phi$.
Therefore, the linear term in the variation of the action~(\ref{cont_change_of_action}) retains the same form, with modified current operators\footnote{Note that the current operators have an ambiguity $J^m_{\tau}\to J^m_{\tau} + \partial_{x}\partial_{y}\Lambda,\,J^m_{xy}\to J^m_{xy} + \partial_{\tau}\Lambda$.}.

Next, we consider the conserved charge operator for the momentum symmetry modulated by an $x$-dependent function $f(x)$,
\begin{align}
    Q^{m}_{x,f}(\tau) = \int dx dy\, f(x) J_{\tau}^{m}(\tau,x,y).
\end{align}
For $\tau_0 < 0 < \tau_1$, we define a shift
\begin{align}
    \alpha_{f}(\tau,x,y) = \alpha_0 f(x) \Theta(\tau_1-\tau) \Theta(\tau-\tau_0).
\end{align}
Eq.~(\ref{cont_change_of_action}) implies
\begin{align}
    Q^{m}_{x,f}(\tau_1) - Q^{m}_{x,f}(\tau_0) = \frac{1}{i}\left.\frac{d}{d\alpha_0} S_0[\phi+\alpha_f]\right|_{\alpha_0=0}.
\end{align}
If we insert a vortex operator $e^{i\phi^{xy}(0,0,0)}$, Eq.~(\ref{cont_flatness}) is violated as
\begin{align}
    \partial_{\tau}(\partial_{x}\partial_{y}\phi)(\tau,x,y) - \partial_{x}\partial_{y}(\partial_{\tau}\phi)(\tau,x,y) = 2\pi \delta(\tau)\delta(x)\delta(y).
\end{align}
Repeating the analysis around Eq.~(\ref{lat_bulk_Witten}), and using Eq.~(\ref{change_of_bulk_theta}), we obtain the Witten effect
\begin{align}
    \Braket{Q_{x,f}^{m}(\tau_1)e^{i\phi^{xy}(0,0,0)}\cdots} - \Braket{e^{i\phi^{xy}(0,0,0)}Q_{x,f}^{m}(\tau_0)\cdots} = -\frac{\btheta}{\pi} f(0) \Braket{e^{i\phi^{xy}(0,0,0)}\cdots},
\end{align}
which is consistent with the lattice result~(\ref{lat_bulk_Witten}).

\subsection{Foliated theta term}
\label{sec_cont_foliated}

We study the foliated theta term in the continuum $\phi$-theory.
The leading-order contribution of Eq.~(\ref{lat_foliated_theta}) in $a_x$ is given by
\begin{align}
    \label{cont_foliated_theta}
    \sfol[\ftheta;\phi] = -\frac{i\lambda^{x}}{(2\pi)^2} \int d\tau dx dy\,\ftheta(x) \left\{\partial_{x}(\partial_{\tau}\phi)\,\partial_{x}(\partial_{x}\partial_{y}\phi) - (\partial_{x}\partial_{y}\phi)\,\partial_{x}^2(\partial_{\tau}\phi)\right\},
\end{align}
where $\lambda^x = a_x^2$.
The parameter $\ftheta(x)$ is a real-valued function depending on the $x$-coordinate.

For the standard compact boson, in which $\partial_{\tau}\phi$, $\partial_{x}\phi$, and $\partial_{y}\phi$ are all well-defined, 
the topological term becomes trivial:
\begin{align}
    \sfol[\ftheta;\phi] = -\frac{i\lambda^{x}}{(2\pi)^2} \int dx\, \ftheta(x) \int d\tau dy\, \left\{\partial_{y}(\partial_{x}\partial_{\tau}\phi \partial_{x}^{2}\phi) - \partial_{\tau}(\partial_{x}\partial_{y}\phi \partial_{x}^2 \phi)\right\} = 0.
\end{align}
In contrast, in the $\phi$-theory, which admits discontinuous field configurations, 
$\partial_{x}^{2}\phi$ is no longer well-defined, and the theta term can become nontrivial.

We again consider the configuration~(\ref{cont_sample_config}).
For this configuration, we find
\begin{align}
    \begin{split}
    \sfol[\ftheta;\phi] = &-i\lambda^{x} \left\{ 2\int dx \, \ftheta(x) (\delta(x-x_0) - \delta(x-(x_0+r_x))) \partial_{x}\delta(x-x_1)\right. \\
    &+ \left. \int dx\, \ftheta{}^{'}(x) \delta(x-x_1) (\delta(x-x_0) - \delta(x-(x_0+r_x)))\right\} \\
    &= +i\lambda^{x} \left[\left(2\ftheta(x)\,\delta'(x-x_1) + \ftheta{}^{'}(x)\,\delta(x-x_1)\right) \right]_{x=x_0}^{x=x_0+r_x},
    \end{split}
\end{align}
which is nonvanishing when $x_1 = x_0$ or $x_0 + r_x$.

Under the shift transformation, we have
\begin{align}
    \begin{split}
        \label{change_of_foliated_theta}
        &\sfol[\ftheta; \phi + \alpha] - \sfol[\ftheta; \phi] \\
        =& \frac{i\lambda^{x}}{(2\pi)^2} \int d\tau dx dy\,\ftheta(x) \left\{\partial_{x} \alpha\, \partial_{x} (\partial_{\tau}(\partial_x\partial_y\phi)-\partial_{x}\partial_{y}(\partial_{\tau}\phi)) - \partial_{x}^2 \alpha\, (\partial_{\tau}(\partial_x\partial_y\phi)-\partial_{x}\partial_{y}(\partial_{\tau}\phi))\right\}.
    \end{split}
\end{align}
In the absence of winding vortices, this expression vanishes due to Eq.~(\ref{cont_flatness}).
In this sense, $\sfol$ is a topological term and does not affect the classical equations of motion.
While preserving this property, one can choose the theta parameter $\ftheta(x)$ to be an arbitrary smooth function.

\paragraph{Witten effect}

Using Eq.~(\ref{change_of_foliated_theta}), we obtain the Witten effect
\begin{align}
    \begin{split}
    &\Braket{Q_{x,f}^{m}(\tau_1)e^{i\phi^{xy}(0,0,0)}\cdots} - \Braket{e^{i\phi^{xy}(0,0,0)}Q_{x,f}^{m}(\tau_0)\cdots}\\
    = & - \frac{\lambda^{x}}{2\pi} \int d\tau dx dy\, \ftheta(x) \left\{f'(x)\, \partial_{x}(\delta(\tau)\delta(x)\delta(y)) - f''(x) \, \delta(\tau)\delta(x)\delta(y) \right\} \Braket{e^{i\phi^{xy}(0,0,0)}\cdots}\\
    = & \frac{\lambda^x}{2\pi} \left(\ftheta{}^{'}(0) f'(0) + 2\ftheta(0)f''(0)\right) \Braket{e^{i\phi^{xy}(0,0,0)}\cdots},
    \end{split}
\end{align}
which is consistent with the leading-order result of its lattice counterpart~(\ref{lat_foliated_Witten_shifted}).

\section{Conclusion and outlook}
\label{section_conclusion}
In this paper, we have studied exotic theta terms in the 2+1d $\phi$-theory.
We introduced two types of theta terms: a bulk theta term and a foliated theta term.
The bulk theta term can be understood as a fractonic analogue of the cup product,
while the foliated theta term is characterized by a theta parameter that can depend on spatial coordinates such as $x$ or $y$.
Both theta terms lead to generalized Witten effects.
We demonstrated these features from both lattice and continuum perspectives.

Let us conclude by commenting on several directions for future work.
One interesting direction is to construct lattice realizations of theta terms in $3+1$d tensor gauge theories, such as those studied in~\cite{Pretko:2017xar}.
Another important problem is to achieve a systematic classification of exotic theta terms in the $\phi$-theory.
It is also of interest to further investigate the properties of the fractonic cup product.

\section*{Acknowledgements}
This work was supported by JST SPRING, Grant Number JPMJSP2108.

\appendix

\section{Topological charge for admissible field configurations in the original XY-plaquette model}
\label{Appendix_topological_charge}

In this appendix, we formulate the topological charge~(\ref{cont_topological_charge}) in the original XY-plaquette model.
Instead of using the modified Villain formulation, we impose an admissibility condition on field configurations following L\"uscher~\cite{Luscher:1981zq}.
This condition requires the field configurations to be "smooth" in an appropriate sense and partitions the space of field configurations into topological sectors.
We then define the topological charge in this setting.
We consider both discrete- and continuum-time formulations.

\subsection{Discrete-time formulation}

In the discrete-time formulation, the spacetime lattice is a cubic lattice with periodic boundary conditions.
For each site $n$, we assign a $\mathrm{U}(1)$-valued dynamical variable $e^{i\phi(n)}$.
We consider the action
\begin{align}
    \label{action-discrete-time}
    S_{\mathrm{XY}} = \beta_0 \sum_{n} \left\{1 - \cos \left(\Dtau \phi(n)\right)\right\} + \beta \sum_{n} \left\{1 - \cos \left(\Dx\Dy \phi(n)\right)\right\},
\end{align}
where $\phi(n) \in \mathbb{R}$ is a lift of $e^{i\phi(n)} \in \mathrm{U}(1)$, and the action~(\ref{action-discrete-time}) is independent of the choice of the lift.

When $\beta_0, \beta \gg 1$, the smoothness conditions $\Dtau \phi, \Dx\Dy \phi \sim 0 \mod 2\pi$ are dynamically imposed.
In this limit, field configurations far from these smoothness conditions do not contribute to the physics.
This justifies imposing the smoothness conditions by hand as an approximation.

Precisely, we fix $\varepsilon_{\tau}, \varepsilon_{xy} > 0$ such that $2 \varepsilon_{\tau} + \varepsilon_{xy} < \pi$.
For each site $n$, we impose the admissibility conditions
\begin{align}
    \label{admissibility}
    \begin{split}
    1 - \cos \left(\Dtau \phi(n)\right) &< 2 \sin^2 \frac{\varepsilon_{\tau}}{2},\\
    1 - \cos \left(\Dx\Dy \phi(n)\right) &< 2 \sin^2 \frac{\varepsilon_{xy}}{2}.
    \end{split}
\end{align}
A configuration is called \emph{admissible} if and only if it satisfies the admissibility conditions~(\ref{admissibility}).

We now construct the topological charge for a given admissible configuration $e^{i\phi}$.
We first define $\partial_{\tau}\phi$ and $\partial_{x}\partial_{y}\phi$ by
\begin{align}
    \begin{split}
    e^{i\partial_{\tau} \phi} = e^{i\Dtau \phi},&\quad -\pi < \partial_{\tau} \phi < \pi, \\
    e^{i\partial_{x}\partial_{y} \phi} = e^{i\Dx\Dy \phi},&\quad -\pi < \partial_{x}\partial_{y} \phi < \pi,
    \end{split}
\end{align}
which are independent of the choice of the lift $\phi \in \mathbb{R}$.

We note that $\abs{\partial_{\tau} \phi} < \varepsilon_{\tau}$ and $\abs{\partial_{x}\partial_{y} \phi} < \varepsilon_{xy}$ due to the admissibility conditions.
Then the topological charge is given by
\begin{align}
    Q = \frac{1}{(2\pi)^2} \sum_{n} \left(\partial_{\tau}\phi(n)\, \partial_{x}\partial_{y}\phi(n+\tauhat) + \partial_{x}\partial_{y}\phi(n)\, \partial_{\tau}\phi(n+\xhat+\yhat)\right).
\end{align}

We next show that it is integer-valued under the admissibility conditions.
Let us pick a lift $\phi \in \mathbb{R}$.
We then obtain integer-valued fields $\ell_{\tau}, \ell_{xy}$ satisfying
\begin{align}
    \begin{split}
    \partial_{\tau} \phi &= \Dtau \phi + 2\pi \ell_{\tau},\\
    \partial_{x}\partial_{y} \phi &= \Dx\Dy \phi + 2\pi \ell_{xy}.
    \end{split}
\end{align}
They satisfy
\begin{align}
    \begin{split}
    \abs{\Dtau \ell_{xy} - \Dx\Dy \ell_{\tau}}
    &= \frac{1}{2\pi} \abs{\Dtau (\partial_{x}\partial_{y}\phi) - \Dx\Dy(\partial_{\tau}\phi)}\\
    &\le \frac{1}{2\pi}\abs{\Dtau (\partial_{x}\partial_{y}\phi)} + \frac{1}{2\pi}\abs{\Dx\Dy(\partial_{\tau}\phi)}.
    \end{split}
\end{align}
Using the admissibility conditions, the right-hand side is bounded by
\begin{align}
    < \frac{2 \varepsilon_{xy} + 4 \varepsilon_{\tau}}{2\pi} < 1,
\end{align}
which implies
\begin{align}
    \Dtau \ell_{xy} - \Dx\Dy \ell_{\tau} = 0.
\end{align}

Using this, we have\footnote{This is essentially the same computation as Eq.~(\ref{lat_rewrite_bulk_theta}).}
\begin{align}
    Q = \sum_{n} \left(\ell_{\tau}(n)\, \ell_{xy}(n+\tauhat) + \ell_{xy}(n)\, \ell_{\tau}(n+\xhat+\yhat)\right),
\end{align}
which implies that $Q \in \mathbb{Z}$.

\subsection{Continuum-time formulation}

We next construct the topological charge in the continuum-time formulation, which corresponds to the path-integral formulation of the model~(\ref{original_hamiltonian}).
In this formulation, a field configuration is given by a $\mathrm{U}(1)$-valued smooth function $e^{i\phi(n;\tau)}$ assigned to each site $n$ on a two-dimensional square lattice with periodic boundary conditions.
Here, $\tau$ is a time coordinate on $S^1$ with the identification $\tau \sim \tau + \ell_{\tau}$.
Although $\phi(n;\tau)$ itself is not globally well-defined, $\partial_{\tau}\phi(n;\tau)$ is well-defined, since it is independent of the choice of the lift $\phi(n;\tau) \in \mathbb{R}$ of $e^{i\phi(n;\tau)}\in\mathrm{U}(1)$.

The Hamiltonian~(\ref{original_hamiltonian}) with $U \ll K$ energetically enforces smoothness along the spatial directions, namely $\Dx\Dy\phi(n;\tau) \sim 0 \mod 2\pi$.
This justifies imposing an admissibility condition by hand as an approximation.
In the continuum-time formulation, the admissibility condition is given by
\begin{align}
    \label{admissibility2}
    e^{i\Dx\Dy\phi(n;\tau)} \neq -1,
\end{align}
for any $n$ and $\tau$.
We define $\partial_{x}\partial_{y}\phi(n;\tau)$ by
\begin{align}
    e^{i\partial_{x}\partial_{y}\phi(n;\tau)} = e^{i\Dx\Dy\phi(n;\tau)},\quad -\pi < \partial_{x}\partial_{y}\phi(n;\tau) < \pi.
\end{align}

We now consider the topological charge
\begin{align}
    Q = \frac{1}{(2\pi)^2} \int_{0}^{\ell_{\tau}} d\tau\, \sum_{n} \left\{\partial_{\tau}\phi(n;\tau)\, \partial_{x}\partial_{y}\phi(n;\tau) + \partial_{x}\partial_{y}\phi(n;\tau)\, \partial_{\tau}\phi(n+\xhat+\yhat;\tau)\right\}.
\end{align}

To show that $Q \in \mathbb{Z}$, we divide the temporal circle $S^1$ into small segments $I_{k} = [\tau_{k},\tau_{k+1}]$, with $k = 0,1,\dots,N-1$.
We set $\tau_0 = 0$ and $\tau_{N} = \ell_{\tau}$, which are identified.
On each segment $I_k$, we choose a lift $\phi_k(n;\tau) \in \mathbb{R}$.
Two adjacent segments are related by
\begin{align}
    \label{tiger}
    \phi_{k-1}(n;\tau_{k}) - \phi_{k}(n;\tau_{k}) = 2\pi m_{k}(n),
\end{align}
where $m_{k}(n) \in \mathbb{Z}$ for each site $n$ and $k = 0,1,\dots,N-1$.\footnote{Here we identify $\phi_{-1}$ with $\phi_{N-1}$.}

For $\tau \in I_{k}$, we have
\begin{align}
    \partial_{x}\partial_{y} \phi(n;\tau) = \Dx\Dy\phi_{k}(n;\tau) + 2\pi \ell_{\tau,k}(n),
\end{align}
with $\ell_{\tau,k}(n) \in \mathbb{Z}$.
We note that $\ell_{\tau,k}(n)$ is independent of $\tau \in I_k$ due to the admissibility condition~(\ref{admissibility2}).

Using Eq.~(\ref{tiger}), we obtain
\begin{align}
    \ell_{\tau,k-1}(n) - \ell_{\tau,k}(n) = -\Dx\Dy m_{k}(n).
\end{align}

Using these relations, we compute
\begin{align}
    \begin{split}
    Q =& \frac{1}{(2\pi)^2}\sum_{n}\sum_{k=0}^{N-1} \int_{I_{k}} d\tau \left\{\partial_{\tau}\phi_{k}(n;\tau)\left(\Dx\Dy\phi_{k}(n;\tau) + 2\pi \ell_{\tau,k}(n)\right) \right.\\
    & \left. + \left(\Dx\Dy\phi_{k}(n;\tau) + 2\pi \ell_{\tau,k}(n)\right) \partial_{\tau}\phi_{k}(n+\xhat+\yhat;\tau)\right\}\\
    =& \frac{1}{(2\pi)^2}\sum_{n}\sum_{k=0}^{N-1} \int_{I_{k}} d\tau\, \partial_{\tau}\Bigl[\phi_{k}(n;\tau)\bigl(\Dx\Dy\phi_{k}(n;\tau)+2\pi \ell_{\tau,k}(n)\bigr) \\
    & \qquad\qquad\qquad\quad + 2\pi \ell_{\tau,k}(n)\phi_{k}(n+\xhat+\yhat;\tau)\Bigr] \\
    =& \frac{1}{2\pi} \sum_{n}\sum_{k=0}^{N-1} \left\{m_{k}(n) \bigl(\Dx\Dy\phi_{k}(n;\tau_k)+2\pi \ell_{\tau,k}(n)\bigr) \right.\\
    & \left. + \ell_{\tau,k-1}(n)\phi_{k-1}(n+\xhat+\yhat;\tau_k) - \ell_{\tau,k}(n)\phi_{k}(n+\xhat+\yhat;\tau_k)\right\}\\
    =& \sum_{n} \sum_{k=0}^{N-1} \left\{m_{k}(n) \ell_{\tau,k}(n) + \ell_{\tau,k}(n)m_{k}(n+\xhat+\yhat) - \Dx\Dy m_{k}(n)\, m_{k}(n+\xhat+\yhat)\right\}\\
    =& \sum_{n} \sum_{k=0}^{N-1} \left\{\ell_{\tau,k}(n) m_{k+1}(n) + \ell_{\tau,k}(n)m_{k}(n+\xhat+\yhat) \right\},
    \end{split}
\end{align}
which implies that $Q \in \mathbb{Z}$.

\section{Proof of $\frac{1}{2\pi^2}\int_{T^3} d\tau dx dy \, \partial_{\tau}\phi \, \partial_{x}\partial_{y}\phi \in \mathbb{Z}$}
\label{appendix_quantization_of_topological_charge}
The goal of this appendix is to show that
\begin{align}
    Q = \frac{1}{2\pi^2}\int_{T^3} d\tau\, dx\, dy \, \partial_{\tau}\phi \, \partial_{x}\partial_{y}\phi
\end{align}
takes integer values.

Let us first clarify the setup.
We consider $\phi$ defined on a three-dimensional torus with the identifications $\tau \sim \tau + \ell_{\tau}$, $x \sim x + \ell_{x}$, and $y \sim y + \ell_{y}$.
As explained in Section~\ref{review_of_continuum}, the fundamental field $\phi$ is not simply a real-valued function.
To specify a field configuration, one must divide spacetime into small patches and assign a local expression $\phi_{\alpha}$ on each patch $\alpha$, together with appropriate transition functions between overlapping patches.
Given an open cover of $T^3$, one can, by refining the cover, reorganize it into a simpler form described below, which we adopt in the following.

We consider a set $\tilde{\Lambda}$ of reference points defined by $\tilde{\Lambda} = \{(\tau_i, x_j, y_k)\}_{(i,j,k)\in \Lambda}$,
where $\Lambda = \{(i,j,k) \mid i,j,k \in \mathbb{Z},\ 0 \le i < I,\ 0 \le j < J,\ 0 \le k < K\}$,
and $0 = \tau_0 \le \tau_1 \le \cdots \le \tau_I = \ell_{\tau}, \quad
    0 = x_0 \le x_1 \le \cdots \le x_J = \ell_x, \quad
    0 = y_0 \le y_1 \le \cdots \le y_K = \ell_y$.
The set $\Lambda$ can be regarded as the sites of a cubic lattice embedded in the continuum spacetime.
We then partition spacetime into rectangular blocks accordingly.

For each $n = (i,j,k) \in \Lambda$, we assign a block
\begin{align}
    c(n) = \{(\tau,x,y) \mid \tau_i \le \tau \le \tau_{i+1},\ x_j \le x \le x_{j+1},\ y_k \le y \le y_{k+1}\}.
\end{align}
For convenience, we also define the following lower-dimensional cells:
\begin{align}
    p_{\tau x}(n) &= \{(\tau,x,y_k) \mid \tau_i \le \tau \le \tau_{i+1},\ x_j \le x \le x_{j+1}\},\\
    p_{\tau y}(n) &= \{(\tau,x_j,y) \mid \tau_i \le \tau \le \tau_{i+1},\ y_k \le y \le y_{k+1}\},\\
    p_{xy}(n) &= \{(\tau_i,x,y) \mid x_j \le x \le x_{j+1},\ y_k \le y \le y_{k+1}\},\\
    \ell_{\tau}(n) &= \{(\tau,x_j,y_k) \mid \tau_i \le \tau \le \tau_{i+1}\},\\
    \ell_{x}(n) &= \{(\tau_i,x,y_k) \mid x_j \le x \le x_{j+1}\},\\
    \ell_{y}(n) &= \{(\tau_i,x_j,y) \mid y_k \le y \le y_{k+1}\},\\
    s(n) &= (\tau_i,x_j,y_k).
\end{align}
Using a constant $0 < \varepsilon < \tfrac{1}{2}$, we define a local patch for each $n \in \Lambda$ by
\begin{align}
    U(n) = \{(\tau,x,y) \mid \tau_i - \varepsilon \le \tau \le \tau_{i+1} + \varepsilon,\ x_j - \varepsilon \le x \le x_{j+1} + \varepsilon,\ y_k - \varepsilon \le y \le y_{k+1} + \varepsilon\},
\end{align}
which is slightly larger than $c(n)$.
On each patch $U(n)$, the field $\phi$ is described by a real-valued function $\phi_n$.

For $\mu = \tau, x, y$, the overlap $U(n) \cap U(n-\hat{\mu})$ is nonempty, on which the local expressions of $\phi$ are related by
\begin{align}
    \phi_{n-\hat{\mu}}(\tau,x,y) = \phi_n(\tau,x,y) + 2\pi\, m_{n,\mu}(x,y),
\end{align}
where the transition function is given by
\begin{align}
    \label{transition_func_decomposition}
    m_{n,\mu}(x,y) = m^x_{n,\mu}(x) + m^y_{n,\mu}(y),
\end{align}
with integer-valued, piecewise constant functions $m^x_{n,\mu}(x)$ and $m^y_{n,\mu}(y)$.

These transition functions satisfy the cocycle condition
\begin{align}
    \label{cocycle_cond_not_xy}
    m_{n,\mu}(x,y) + m_{n-\hat{\mu},\nu}(x,y) = m_{n,\nu}(x,y) + m_{n-\hat{\nu},\mu}(x,y),
\end{align}
for $\mu,\nu = \tau, x, y$ with $\mu \neq \nu$.
Using Eq.~(\ref{transition_func_decomposition}), this can be rewritten as
\begin{align}
    \label{cocycle_cond_xy}
    \begin{split}
    &m^x_{n,\mu}(x) + m^x_{n-\hat{\mu},\nu}(x) - m^x_{n,\nu}(x) - m^x_{n-\hat{\nu},\mu}(x) \\
    =& -\bigl(m^y_{n,\mu}(y) + m^y_{n-\hat{\mu},\nu}(y) - m^y_{n,\nu}(y) - m^y_{n-\hat{\nu},\mu}(y)\bigr).
    \end{split}
\end{align}

We assume that the number of discontinuities of $m_{n,\mu}(x,y)$ is finite.
We further assume that no discontinuity of any transition function coincides with the boundary of any block.
This can always be achieved by an appropriate choice of the set $\tilde{\Lambda}$ defining the open cover of spacetime.
We note that the following computations rely on the Leibniz rule, in particular the one for the squared step function~(\ref{Leibniz}), which in turn implies Eq.~(\ref{our_prescription}).

Under this setup, we prove that $Q \in \mathbb{Z}$.
First, note that the integrand can be locally expressed as
\begin{align}
    2\,\partial_{\tau}\phi\, \partial_{x}\partial_{y}\phi
    = \partial_{\tau}(\phi\,\partial_{x}\partial_{y}\phi)
    - \partial_{x}(\phi\,\partial_{\tau}\partial_{y}\phi)
    - \partial_{y}(\phi\,\partial_{\tau}\partial_{x}\phi)
    + \partial_{x}\partial_{y}(\phi\,\partial_{\tau}\phi).
\end{align}
Therefore, we obtain
\begin{align}
    Q = \frac{1}{(2\pi)^2} \left( I_{xy} - I_{\tau x} - I_{\tau y} + I_{\tau} \right),
\end{align}
where
\begin{align}
    I_{xy} &= \sum_{n} \int_{c(n)} d\tau\, dx\, dy\, \partial_{\tau}\bigl(\phi_{n}\,\partial_{x}\partial_{y}\phi_{n}\bigr),\\
    I_{\tau x} &= \sum_{n} \int_{c(n)} d\tau\, dx\, dy\, \partial_{y}\bigl(\phi_{n}\,\partial_{\tau}\partial_{x}\phi_{n}\bigr),\\
    I_{\tau y} &= \sum_{n} \int_{c(n)} d\tau\, dx\, dy\, \partial_{x}\bigl(\phi_{n}\,\partial_{\tau}\partial_{y}\phi_{n}\bigr),\\
    I_{\tau} &= \sum_{n} \int_{c(n)} d\tau\, dx\, dy\, \partial_{x}\partial_{y}\bigl(\phi_{n}\,\partial_{\tau}\phi_{n}\bigr).
\end{align}

We compute them separately.
Firstly, we have
\begin{align}
    \begin{split}
    I_{xy} =& \sum_{n} \int_{p_{xy}(n+\tauhat) - p_{xy}(n)} \phi_{n}\partial_{x}\partial_{y}\phi_{n}\\
    =& \sum_{n}\int_{p_{xy}(n)} (\phi_{n-\tauhat}\partial_{x}\partial_{y}\phi_{n-\tauhat} - \phi_{n}\partial_{x}\partial_{y}\phi_{n})\\
    =& \sum_{n}\int_{p_{xy}(n)} (\phi_{n-\tauhat}-\phi_{n})\partial_{x}\partial_{y}\phi_{n}\\
    =& 2\pi \sum_{n} \int_{p_{xy}(n)} (m^x_{n,\tau} + m^y_{n,\tau}) \partial_{x}\partial_{y}\phi_{n}\\
    =& 2\pi \sum_{n} \int_{p_{xy}(n)} \left\{\partial_{x}(m^y_{n,\tau}\partial_{y}\phi_{n}) + \partial_{y}(m^x_{n,\tau}\partial_{x}\phi_{n})\right\}\\
    =& 2\pi \sum_{n} \left(\int_{\ell_{y}(n+\xhat)-\ell_{y}(n)} m^y_{n,\tau}\partial_{y}\phi_{n} + \int_{\ell_{x}(n+\yhat)-\ell_{x}(n)} m^x_{n,\tau}\partial_{x}\phi_{n}\right) \\
    =& 2\pi \sum_{n} \left\{\int_{\ell_{y}(n)} \left(m^y_{n-\xhat,\tau}\partial_{y}\phi_{n-\xhat} - m^{y}_{n,\tau}\partial_{y}\phi_{n}\right)
    + \int_{\ell_{x}(n)}\left(m^x_{n-\yhat,\tau}\partial_{x}\phi_{n-\yhat} - m^x_{n,\tau} \partial_{x}\phi_{n} \right) \right\}\\
    =& 2\pi \sum_{n} \left\{\int_{\ell_{y}(n)} (m^y_{n-\xhat,\tau}-m^y_{n,\tau})\partial_{y}\phi_{n} + \int_{\ell_{x}(n)} (m^x_{n-\yhat,\tau}-m^x_{n,\tau})\partial_{x}\phi_{n}\right.\\
    &+ \left. 2\pi \int_{\ell_{y}(n)} m^y_{n-\xhat,\tau}\partial_{y} m^y_{n,x} + 2\pi \int_{\ell_{x}(n)} m^x_{n-\yhat,\tau}\partial_{x}m^{x}_{n,y} \right\}.
    \end{split}
\end{align}
The second contribution is given by
\begin{align}
    \begin{split}
    I_{\tau x} =& \sum_{n} \int_{p_{\tau x}(n+\yhat) - p_{\tau x}(n)} \phi_{n}\partial_{\tau}\partial_{x}\phi_{n}\\
    =& \sum_{n} \int_{p_{\tau x}(n)} (\phi_{n-\yhat}\partial_{\tau}\partial_{x}\phi_{n-\yhat} - \phi_{n}\partial_{\tau}\partial_{x}\phi_{n}) \\
    =& 2\pi \sum_{n} \int_{p_{\tau x}(n)} (m^x_{n,y}+m^y_{n,y}) \partial_{\tau}\partial_{x}\phi_{n} \\
    =& 2\pi \sum_{n} \int_{p_{\tau x}(n)} \left\{ \partial_{\tau}\partial_{x}(m^y_{n,y}\phi_{n}) + \partial_{\tau} (m^x_{n,y}\partial_{x}\phi_{n})\right\} \\
    =& 2\pi \sum_{n}  \left\{\int_{\ell_{x}(n)} \left(m^x_{n-\tauhat,y} \partial_{x} \phi_{n-\tauhat} - m^x_{n,y}\partial_{x}\phi_{n} \right)\right. \\
    &+ \left. m^y_{n,y} \phi_{n}(s(n+\tauhat+\xhat)) - m^y_{n,y} \phi_{n}(s(n+\tauhat)) - m^y_{n,y} \phi_{n}(s(n+\xhat)) + m^y_{n,y} \phi_{n}(s(n)) \right\} \\
    =& 2\pi \sum_{n} \left\{ \int_{\ell_{x}(n)} (m^x_{n-\tauhat,y} - m^x_{n,y})\partial_{x}\phi_{n}
    + 2\pi \int_{\ell_{x}(n)} m^x_{n-\tauhat,y}\partial_{x}m^x_{n,\tau} \right.\\
    & \left. +\left.(m^y_{n-\tauhat-\xhat,y}\phi_{n-\tauhat-\xhat}-m^y_{n-\tauhat,y}\phi_{n-\tauhat}-m^y_{n-\xhat,y}\phi_{n-\xhat}+m^y_{n,y}\phi_{n})\right|_{s(n)}\right\}.
    \end{split}
\end{align}
Similarly, we have
\begin{align}
    \begin{split}
    I_{\tau y}=& 2\pi \sum_{n} \left\{ \int_{\ell_{y}(n)} (m^y_{n-\tauhat,x} - m^y_{n,x})\partial_{y}\phi_{n}
    + 2\pi \int_{\ell_{y}(n)} m^y_{n-\tauhat,x}\partial_{y}m^y_{n,\tau} \right. \\
    & \left. +\left.(m^x_{n-\tauhat-\yhat,x}\phi_{n-\tauhat-\yhat}-m^x_{n-\tauhat,x}\phi_{n-\tauhat}-m^x_{n-\yhat,x}\phi_{n-\yhat}+m^x_{n,x}\phi_{n})\right|_{s(n)}\right\}.
    \end{split}
\end{align}
Combining these three terms and using Eq.~(\ref{cocycle_cond_xy}), we obtain
\begin{align}
    \begin{split}
    &\frac{I_{xy} - I_{\tau x} - I_{\tau y}}{2\pi}\\
    =& \sum_{n} \left\{\int_{\ell_{x}(n)} (m^x_{n-\yhat,\tau}-m^x_{n,\tau}-m^x_{n-\tauhat,y} + m^x_{n,y})\partial_{x}\phi_{n}\right.\\
    &+ \int_{\ell_{y}(n)} (m^y_{n-\xhat,\tau}-m^y_{n,\tau}-m^y_{n-\tauhat,x} + m^y_{n,x})\partial_{y}\phi_{n}\\
    &-\left.(m^y_{n-\tauhat-\xhat,y}\phi_{n-\tauhat-\xhat}-m^y_{n-\tauhat,y}\phi_{n-\tauhat}-m^y_{n-\xhat,y}\phi_{n-\xhat}+m^y_{n,y}\phi_{n})\right|_{s(n)}\\
    &-\left.(m^x_{n-\tauhat-\yhat,x}\phi_{n-\tauhat-\yhat}-m^x_{n-\tauhat,x}\phi_{n-\tauhat}-m^x_{n-\yhat,x}\phi_{n-\yhat}+m^x_{n,x}\phi_{n})\right|_{s(n)}\\
    &- 2\pi \int_{\ell_{x}(n)} m^x_{n-\tauhat,y}\partial_{x}m^x_{n,\tau} - 2\pi \int_{\ell_{y}(n)} m^y_{n-\tauhat,x}\partial_{y}m^y_{n,\tau}\\
    &+ \left. 2\pi \int_{\ell_{y}(n)} m^y_{n-\xhat,\tau}\partial_{y} m^y_{n,x} + 2\pi \int_{\ell_{x}(n)} m^x_{n-\yhat,\tau}\partial_{x}m^{x}_{n,y} \right\}\\
    =& \sum_{n} \left\{(m^x_{n-\xhat-\yhat,\tau}-m^x_{n-\xhat,\tau}-m^x_{n-\tauhat-\xhat,y}+m^x_{n-\xhat,y}-m^x_{n-\yhat,\tau}+m^x_{n,\tau}+m^x_{n-\tauhat,y}-m^x_{n,y})\phi_{n}\right|_{s(n)}\\
    & + \left. 2\pi (m^x_{n-\xhat-\yhat,\tau}-m^x_{n-\xhat,\tau}-m^x_{n-\tauhat-\xhat,y} + m^x_{n-\xhat,y}) m_{n,x}\right|_{s(n)}\\
    & + \left.(m^y_{n-\xhat-\yhat,\tau}-m^y_{n-\yhat,\tau}-m^y_{n-\tauhat-\yhat,x}+m^y_{n-\yhat,x}-m^y_{n-\xhat,\tau}+m^y_{n,\tau}+m^y_{n-\tauhat,x}-m^y_{n,x})\phi_{n}\right|_{s(n)}\\
    & + \left. 2\pi (m^y_{n-\xhat-\yhat,\tau}-m^y_{n-\yhat,\tau}-m^y_{n-\tauhat-\yhat,x} + m^y_{n-\yhat,x}) m_{n,y}\right|_{s(n)}\\
    &-\left.(m^y_{n-\tauhat-\xhat,y}\phi_{n-\tauhat-\xhat}-m^y_{n-\tauhat,y}\phi_{n-\tauhat}-m^y_{n-\xhat,y}\phi_{n-\xhat}+m^y_{n,y}\phi_{n})\right|_{s(n)}\\
    &-\left.(m^x_{n-\tauhat-\yhat,x}\phi_{n-\tauhat-\yhat}-m^x_{n-\tauhat,x}\phi_{n-\tauhat}-m^x_{n-\yhat,x}\phi_{n-\yhat}+m^x_{n,x}\phi_{n})\right|_{s(n)}\\
    &- 2\pi \int_{\ell_{x}(n)} m^x_{n-\tauhat,y}\partial_{x}m^x_{n,\tau} - 2\pi \int_{\ell_{y}(n)} m^y_{n-\tauhat,x}\partial_{y}m^y_{n,\tau}\\
    &+ \left. 2\pi \int_{\ell_{y}(n)} m^y_{n-\xhat,\tau}\partial_{y} m^y_{n,x} + 2\pi \int_{\ell_{x}(n)} m^x_{n-\yhat,\tau}\partial_{x}m^{x}_{n,y} \right\}\\
    =& \sum_{n}\left\{
    \left(m_{n-\xhat-\yhat,\tau}-m_{n-\xhat,\tau}-m_{n-\yhat,\tau}+m_{n-\tauhat,y}+m_{n-\tauhat,x}+m_{n-\xhat,y}\right.\right.\\
    &\left.\left.+m_{n-\yhat,x}-m_{n,x}-m_{n,y}-m_{n-\tauhat-\xhat,y}-m_{n-\tauhat-\yhat,x}+m_{n,\tau}\right)\phi_{n}\right|_{s(n)}\\
    &-\left.(m^y_{n-\tauhat-\xhat,y}(\phi_{n-\tauhat-\xhat}-\phi_{n})-m^y_{n-\tauhat,y}(\phi_{n-\tauhat}-\phi_{n})-m^y_{n-\xhat,y}(\phi_{n-\xhat}-\phi_{n}))\right|_{s(n)}\\
    &-\left.(m^x_{n-\tauhat-\yhat,x}(\phi_{n-\tauhat-\yhat}-\phi_{n})-m^x_{n-\tauhat,x}(\phi_{n-\tauhat}-\phi_{n})-m^x_{n-\yhat,x}(\phi_{n-\yhat}-\phi_{n}))\right|_{s(n)}\\
    & + \left. 2\pi (m^x_{n-\xhat-\yhat,\tau}-m^x_{n-\xhat,\tau}-m^x_{n-\tauhat-\xhat,y} + m^x_{n-\xhat,y}) m_{n,x}\right|_{s(n)}\\
    & + \left. 2\pi (m^y_{n-\xhat-\yhat,\tau}-m^y_{n-\yhat,\tau}-m^y_{n-\tauhat-\yhat,x} + m^y_{n-\yhat,x}) m_{n,y}\right|_{s(n)}\\
    &- 2\pi \int_{\ell_{x}(n)} m^x_{n-\tauhat,y}\partial_{x}m^x_{n,\tau} - 2\pi \int_{\ell_{y}(n)} m^y_{n-\tauhat,x}\partial_{y}m^y_{n,\tau}\\
    &+ \left. 2\pi \int_{\ell_{y}(n)} m^y_{n-\xhat,\tau}\partial_{y} m^y_{n,x} + 2\pi \int_{\ell_{x}(n)} m^x_{n-\yhat,\tau}\partial_{x}m^{x}_{n,y} \right\}.
    \end{split}
\end{align}
On the other hand,
\begin{align}
    \begin{split}
    I_{\tau} =& \sum_{n}\int_{\ell_{\tau}(n+\xhat+\yhat)-\ell_{\tau}(n+\xhat)-\ell_{\tau}(n+\yhat)+\ell_{\tau}(n)} \phi_{n}\partial_{\tau}\phi_{n}\\
    =& \sum_{n} \int_{\ell_{\tau}(n)} (\phi_{n-\xhat-\yhat}\partial_{\tau}\phi_{n-\xhat-\yhat} - \phi_{n-\xhat}\partial_{\tau}\phi_{n-\xhat} - \phi_{n-\yhat}\partial_{\tau}\phi_{n-\yhat} + \phi_{n}\partial_{\tau}\phi_{n})\\
    =& \sum_{n} \int_{\ell_{\tau}(n)} \partial_{\tau}\left\{(\phi_{n-\xhat-\yhat}-\phi_{n-\xhat}-\phi_{n-\yhat}+\phi_{n})\phi_{n}\right\}\\
    =& \sum_{n} \left[ \left.\{(\phi_{n-\tauhat-\xhat-\yhat}-\phi_{n-\tauhat-\xhat}-\phi_{n-\tauhat-\yhat}+\phi_{n-\tauhat})-(\phi_{n-\xhat-\yhat}-\phi_{n-\xhat}-\phi_{n-\yhat}+\phi_{n})\}\phi_{n}\right|_{s(n)} \right.\\
    & \left. + 2\pi \left.(\phi_{n-\tauhat-\xhat-\yhat}-\phi_{n-\tauhat-\xhat}-\phi_{n-\tauhat-\yhat}+\phi_{n-\tauhat})m_{n,\tau}\right|_{s(n)}\right]\\
    =& 2\pi \sum_{n} \left.\left\{(m_{n-\xhat-\yhat,\tau}-m_{n-\xhat,\tau}-m_{n-\yhat,\tau}+m_{n,\tau})\phi_{n}
    + (\phi_{n-\tauhat-\xhat-\yhat}-\phi_{n-\tauhat-\xhat}-\phi_{n-\tauhat-\yhat}+\phi_{n-\tauhat})m_{n,\tau}\right\}\right|_{s(n)}.
    \end{split}
\end{align}

After remarkable cancellations due to the cocycle condition~(\ref{cocycle_cond_not_xy}), we have
\begin{align}
    \begin{split}
    Q =& \sum_{n} \left\{\left. \frac{1}{2\pi}(\phi_{n-\tauhat-\xhat-\yhat}-\phi_{n-\tauhat-\xhat}-\phi_{n-\tauhat-\yhat}+\phi_{n-\tauhat})m_{n,\tau}\right|_{s(n)}\right.\\
    &-\frac{1}{2\pi}\left.(m^y_{n-\tauhat-\xhat,y}(\phi_{n-\tauhat-\xhat}-\phi_{n})-m^y_{n-\tauhat,y}(\phi_{n-\tauhat}-\phi_{n})-m^y_{n-\xhat,y}(\phi_{n-\xhat}-\phi_{n}))\right|_{s(n)}\\
    &-\frac{1}{2\pi}\left.(m^x_{n-\tauhat-\yhat,x}(\phi_{n-\tauhat-\yhat}-\phi_{n})-m^x_{n-\tauhat,x}(\phi_{n-\tauhat}-\phi_{n})-m^x_{n-\yhat,x}(\phi_{n-\yhat}-\phi_{n}))\right|_{s(n)}\\
    & + \left. (m^x_{n-\xhat-\yhat,\tau}-m^x_{n-\xhat,\tau}-m^x_{n-\tauhat-\xhat,y} + m^x_{n-\xhat,y}) m_{n,x}\right|_{s(n)}\\
    & + \left. (m^y_{n-\xhat-\yhat,\tau}-m^y_{n-\yhat,\tau}-m^y_{n-\tauhat-\yhat,x} + m^y_{n-\yhat,x}) m_{n,y}\right|_{s(n)}\\
    &\left. - \int_{\ell_{x}(n)} m^x_{n-\tauhat,y}\partial_{x}m^x_{n,\tau} - \int_{\ell_{y}(n)} m^y_{n-\tauhat,x}\partial_{y}m^y_{n,\tau} + \int_{\ell_{y}(n)} m^y_{n-\xhat,\tau}\partial_{y} m^y_{n,x} + \int_{\ell_{x}(n)} m^x_{n-\yhat,\tau}\partial_{x}m^{x}_{n,y} \right\}.\\
    \end{split}
\end{align}
All terms except for the last line are manifestly integer-valued.
However, the contribution from the last line is more subtle, as its integrality is not manifest due to the prescription~(\ref{our_prescription}).

It remains to show that, for a fixed $c \in \{0,1,\cdots, J-1\}$,
\begin{align}
    \mathcal{I}_{c} = \sum_{i,k} \left(
    \int_{\ell_{x}(i,c,k)} m^x_{(i,c,k-1),\tau}\,\partial_{x}m^{x}_{(i,c,k),y}
    - \int_{\ell_{x}(i,c,k)} m^x_{(i-1,c,k),y}\,\partial_{x}m^x_{(i,c,k),\tau}
    \right) \in \mathbb{Z}.
\end{align}
Note that, for $\mu = \tau, y$ and $x_c < x < x_{c+1}$, the transition function can be written as
\begin{align}
    m_{(i,c,k),\mu}^{x}(x)
    = \sum_{\alpha=0}^{A} W_{(i,k),\mu}^{x,\alpha} \, \Theta(x - X_{\alpha}),
\end{align}
for some $X_0 < x_c < X_1 < X_2 < \cdots < X_A < x_{c+1}$, where $W_{(i,k),\mu}^{x,\alpha} \in \mathbb{Z}$.

Eq.~(\ref{cocycle_cond_xy}) then implies that
\begin{align}
    \begin{split}
    0 = & \partial_{x}\left(m^x_{(i,c,k),\tau}(x) + m^x_{(i-1,c,k),y}(x) - m^x_{(i,c,k),y}(x) - m^x_{(i,c,k-1),\tau}(x)\right) \\
    = & \sum_{\alpha=1}^{A} \left(W^{x,\alpha}_{(i,k),\tau} + W^{x,\alpha}_{(i-1,k),y} - W^{x,\alpha}_{(i,k),y} - W^{x,\alpha}_{(i,k-1),\tau}\right) \delta(x-X_{\alpha}).
    \end{split}
\end{align}
Hence, $W^{x,\alpha}\,(\alpha=1,2,\cdots, A)$ defines a $\mathbb{Z}$-valued $1$-cocycle on a square lattice on a two-dimensional torus.

Using the prescription~(\ref{our_prescription}), we obtain
\begin{align}
    \begin{split}
    \mathcal{I}_{c}
    = & \sum_{i,k}\sum_{\alpha=0}^{A}\sum_{\beta=0}^{A}
    \left(W^{x,\alpha}_{(i,k-1),\tau}W^{x,\beta}_{(i,k),y}-W^{x,\alpha}_{(i-1,k),y}W^{x,\beta}_{(i,k),\tau}\right)
    \int_{\ell_{x}(i,c,k)} dx\, \Theta(x-X_{\alpha}) \delta(x-X_{\beta}) \\
    = & \frac{1}{2}\sum_{i,k}\sum_{\alpha=1}^{A}
    \left(W^{x,\alpha}_{(i,k-1),\tau}W^{x,\alpha}_{(i,k),y}-W^{x,\alpha}_{(i-1,k),y}W^{x,\alpha}_{(i,k),\tau}\right)
    + \text{(integer)}.
    \end{split}
\end{align}
The first term vanishes, since it can be rewritten as
$\sum_{\alpha}\int W^{x,\alpha} \cup W^{x,\alpha}$, which is zero.
Therefore, $\mathcal{I}_{c} \in \mathbb{Z}$.
We conclude that $Q \in \mathbb{Z}$.\footnote{
The above argument might suggest that $\mathcal{I}_{c} \in \mathbb{Z}$ holds regardless of the value of
$\int dx\, \Theta(x-x_{\alpha})\delta(x-x_{\alpha})$, and hence that $Q \in \mathbb{Z}$ holds as well.
However, this is not the case: the prescription~(\ref{our_prescription}) is already implicitly assumed,
since the Leibniz rule~(\ref{Leibniz}) used in the preceding derivation requires it.
}

\bibliographystyle{JHEP}
\bibliography{fractonic-theta.bib}

@article{Seiberg:2020bhn,
    author = "Seiberg, Nathan and Shao, Shu-Heng",
    title = "{Exotic Symmetries, Duality, and Fractons in 2+1-Dimensional Quantum Field Theory}",
    eprint = "2003.10466",
    archivePrefix = "arXiv",
    primaryClass = "cond-mat.str-el",
    doi = "10.21468/SciPostPhys.10.2.027",
    journal = "SciPost Phys.",
    volume = "10",
    number = "2",
    pages = "027",
    year = "2021"
}

@article{Seiberg:2020wsg,
    author = "Seiberg, Nathan and Shao, Shu-Heng",
    title = "{Exotic $U(1)$ Symmetries, Duality, and Fractons in 3+1-Dimensional Quantum Field Theory}",
    eprint = "2004.00015",
    archivePrefix = "arXiv",
    primaryClass = "cond-mat.str-el",
    doi = "10.21468/SciPostPhys.9.4.046",
    journal = "SciPost Phys.",
    volume = "9",
    number = "4",
    pages = "046",
    year = "2020"
}

@article{Seiberg:2020cxy,
    author = "Seiberg, Nathan and Shao, Shu-Heng",
    title = "{Exotic $\mathbb{Z}_N$ symmetries, duality, and fractons in 3+1-dimensional quantum field theory}",
    eprint = "2004.06115",
    archivePrefix = "arXiv",
    primaryClass = "cond-mat.str-el",
    doi = "10.21468/SciPostPhys.10.1.003",
    journal = "SciPost Phys.",
    volume = "10",
    number = "1",
    pages = "003",
    year = "2021"
}

@article{Gorantla:2021svj,
    author = "Gorantla, Pranay and Lam, Ho Tat and Seiberg, Nathan and Shao, Shu-Heng",
    title = "{A modified Villain formulation of fractons and other exotic theories}",
    eprint = "2103.01257",
    archivePrefix = "arXiv",
    primaryClass = "cond-mat.str-el",
    doi = "10.1063/5.0060808",
    journal = "J. Math. Phys.",
    volume = "62",
    number = "10",
    pages = "102301",
    year = "2021"
}

@article{Gorantla:2021bda,
    author = "Gorantla, Pranay and Lam, Ho Tat and Seiberg, Nathan and Shao, Shu-Heng",
    title = "{Low-energy limit of some exotic lattice theories and UV/IR mixing}",
    eprint = "2108.00020",
    archivePrefix = "arXiv",
    primaryClass = "cond-mat.str-el",
    doi = "10.1103/PhysRevB.104.235116",
    journal = "Phys. Rev. B",
    volume = "104",
    number = "23",
    pages = "235116",
    year = "2021"
}

@article{Burnell:2021reh,
    author = "Burnell, Fiona J. and Devakul, Trithep and Gorantla, Pranay and Lam, Ho Tat and Shao, Shu-Heng",
    title = "{Anomaly inflow for subsystem symmetries}",
    eprint = "2110.09529",
    archivePrefix = "arXiv",
    primaryClass = "cond-mat.str-el",
    reportNumber = "MIT-CTP/5336",
    doi = "10.1103/PhysRevB.106.085113",
    journal = "Phys. Rev. B",
    volume = "106",
    number = "8",
    pages = "085113",
    year = "2022"
}

@article{Spieler:2024fby,
    author = "Spieler, Ryan C.",
    title = "{Non-invertible duality interfaces in field theories with exotic symmetries}",
    eprint = "2402.14944",
    archivePrefix = "arXiv",
    primaryClass = "hep-th",
    doi = "10.1007/JHEP06(2024)042",
    journal = "JHEP",
    volume = "06",
    pages = "042",
    year = "2024"
}

@article{Ohmori:2025fuy,
    author = "Ohmori, Kantaro and Shimamura, Shutaro",
    title = "{Gapless foliated-exotic duality}",
    eprint = "2504.10835",
    archivePrefix = "arXiv",
    primaryClass = "cond-mat.str-el",
    doi = "10.1007/JHEP09(2025)049",
    journal = "JHEP",
    volume = "09",
    pages = "049",
    year = "2025"
}

@article{Apruzzi:2025mdl,
    author = "Apruzzi, Fabio and Bedogna, Francesco and Mancani, Salvo",
    title = "{SymTFT construction of gapless exotic-foliated dual models}",
    eprint = "2504.11449",
    archivePrefix = "arXiv",
    primaryClass = "cond-mat.str-el",
    month = "4",
    year = "2025"
}

@article{Bedogna:2026bck,
    author = "Bedogna, Francesco and Mancani, Salvo",
    title = "{The Line, the Strip and the Duality Defect}",
    eprint = "2602.03926",
    archivePrefix = "arXiv",
    primaryClass = "hep-th",
    month = "2",
    year = "2026"
}

@article{Pretko:2017xar,
    author = "Pretko, Michael",
    title = "{Higher-Spin Witten Effect and Two-Dimensional Fracton Phases}",
    eprint = "1707.03838",
    archivePrefix = "arXiv",
    primaryClass = "cond-mat.str-el",
    doi = "10.1103/PhysRevB.96.125151",
    journal = "Phys. Rev. B",
    volume = "96",
    number = "12",
    pages = "125151",
    year = "2017"
}

@article{Nandkishore:2018sel,
    author = "Nandkishore, Rahul M. and Hermele, Michael",
    title = "{Fractons}",
    eprint = "1803.11196",
    archivePrefix = "arXiv",
    primaryClass = "cond-mat.str-el",
    doi = "10.1146/annurev-conmatphys-031218-013604",
    journal = "Ann. Rev. Condensed Matter Phys.",
    volume = "10",
    pages = "295--313",
    year = "2019"
}

@article{Pretko:2020cko,
    author = "Pretko, Michael and Chen, Xie and You, Yizhi",
    title = "{Fracton Phases of Matter}",
    eprint = "2001.01722",
    archivePrefix = "arXiv",
    primaryClass = "cond-mat.str-el",
    doi = "10.1142/S0217751X20300033",
    journal = "Int. J. Mod. Phys. A",
    volume = "35",
    number = "06",
    pages = "2030003",
    year = "2020"
}

@article{Paramekanti:2002iup,
    author = "Paramekanti, Arun and Balents, Leon and Fisher, Matthew P. A.",
    title = "{Ring exchange, the exciton Bose liquid, and bosonization in two dimensions}",
    doi = "10.1103/PhysRevB.66.054526",
    journal = "Phys. Rev. B",
    volume = "66",
    number = "5",
    pages = "054526",
    year = "2002"
}

@article{Abe:2023uan,
    author = "Abe, Motokazu and Morikawa, Okuto and Onoda, Soma and Suzuki, Hiroshi and Tanizaki, Yuya",
    title = "{Magnetic operators in 2D compact scalar field theories on the lattice}",
    eprint = "2304.14815",
    archivePrefix = "arXiv",
    primaryClass = "hep-lat",
    reportNumber = "KYUSHU-HET-260, OU-HET-1185, YITP-23-58",
    doi = "10.1093/ptep/ptad078",
    journal = "PTEP",
    volume = "2023",
    number = "7",
    pages = "073B01",
    year = "2023"
}

@article{Onoda:2025gqa,
    author = "Onoda, Soma",
    title = "{{\textquoteright}t~Hooft Line in 4D U~(1) Lattice Gauge Theory and a Microscopic Description of Dyon{\textquoteright}s Statistics}",
    eprint = "2505.05050",
    archivePrefix = "arXiv",
    primaryClass = "hep-lat",
    reportNumber = "KYUSHU-HET-320",
    doi = "10.1093/ptep/ptaf176",
    journal = "PTEP",
    volume = "2026",
    number = "1",
    pages = "013B04",
    year = "2026"
}

@article{Honda:2024sdz,
    author = "Honda, Yamato and Onoda, Soma and Suzuki, Hiroshi",
    title = "{Action of the Axial U(1) Non-Invertible Symmetry on the {\textquoteright}t~Hooft Line Operator: A Lattice Gauge Theory Study}",
    eprint = "2403.16752",
    archivePrefix = "arXiv",
    primaryClass = "hep-lat",
    reportNumber = "KYUSHU-HET-286",
    doi = "10.1093/ptep/ptae093",
    journal = "PTEP",
    volume = "2024",
    number = "7",
    pages = "073B04",
    year = "2024"
}

@article{Honda:2024xmk,
    author = "Honda, Yamato and Onoda, Soma and Suzuki, Hiroshi",
    title = "{Action of the Axial U(1) Noninvertible Symmetry on the {\textquoteright}t Hooft Line Operator: A Simple Argument}",
    eprint = "2405.07669",
    archivePrefix = "arXiv",
    primaryClass = "hep-lat",
    reportNumber = "KYUSHU-HET-290",
    doi = "10.1093/ptep/ptae167",
    journal = "PTEP",
    volume = "2024",
    number = "11",
    pages = "113B02",
    year = "2024"
}

@article{Katayama:2025pmz,
    author = "Katayama, Nagare and Tanizaki, Yuya",
    title = "{2d Cardy-Rabinovici model with the modified Villain lattice: exact dualities and symmetries}",
    eprint = "2505.19412",
    archivePrefix = "arXiv",
    primaryClass = "hep-th",
    reportNumber = "YITP-25-78",
    doi = "10.1007/JHEP11(2025)004",
    journal = "JHEP",
    volume = "11",
    pages = "004",
    year = "2025"
}

@article{Jacobson:2023cmr,
    author = "Jacobson, Theodore and Sulejmanpasic, Tin",
    title = "{Modified Villain formulation of Abelian Chern-Simons theory}",
    eprint = "2303.06160",
    archivePrefix = "arXiv",
    primaryClass = "hep-th",
    doi = "10.1103/PhysRevD.107.125017",
    journal = "Phys. Rev. D",
    volume = "107",
    number = "12",
    pages = "125017",
    year = "2023"
}

@article{Jacobson:2024hov,
    author = "Jacobson, Theodore and Sulejmanpasic, Tin",
    title = "{Canonical quantization of lattice Chern-Simons theory}",
    eprint = "2401.09597",
    archivePrefix = "arXiv",
    primaryClass = "hep-th",
    doi = "10.1007/JHEP11(2024)087",
    journal = "JHEP",
    volume = "11",
    pages = "087",
    year = "2024"
}

@article{Jacobson:2024muj,
    author = "Jacobson, Theodore",
    title = "{Gauging C on the lattice}",
    eprint = "2406.12075",
    archivePrefix = "arXiv",
    primaryClass = "hep-th",
    doi = "10.1007/JHEP05(2025)138",
    journal = "JHEP",
    volume = "05",
    pages = "138",
    year = "2025"
}

@article{Sulejmanpasic:2019ytl,
    author = "Sulejmanpasic, Tin and Gattringer, Christof",
    title = "{Abelian gauge theories on the lattice: $\theta$-Terms and compact gauge theory with(out) monopoles}",
    eprint = "1901.02637",
    archivePrefix = "arXiv",
    primaryClass = "hep-lat",
    doi = "10.1016/j.nuclphysb.2019.114616",
    journal = "Nucl. Phys. B",
    volume = "943",
    pages = "114616",
    year = "2019"
}

@article{Anosova:2022cjm,
    author = "Anosova, Mariia and Gattringer, Christof and Sulejmanpasic, Tin",
    title = "{Self-dual U(1) lattice field theory with a {\ensuremath{\theta}}-term}",
    eprint = "2201.09468",
    archivePrefix = "arXiv",
    primaryClass = "hep-lat",
    doi = "10.1007/JHEP04(2022)120",
    journal = "JHEP",
    volume = "04",
    pages = "120",
    year = "2022"
}

@article{Anosova:2022yqx,
    author = "Anosova, Mariia and Gattringer, Christof and Iqbal, Nabil and Sulejmanpasic, Tin",
    title = "{Phase structure of self-dual lattice gauge theories in 4d}",
    eprint = "2203.14774",
    archivePrefix = "arXiv",
    primaryClass = "hep-th",
    doi = "10.1007/JHEP06(2022)149",
    journal = "JHEP",
    volume = "06",
    pages = "149",
    year = "2022"
}

@article{Fazza:2022fss,
    author = "Fazza, Lucca and Sulejmanpasic, Tin",
    title = "{Lattice quantum Villain Hamiltonians: compact scalars, U(1) gauge theories, fracton models and quantum Ising model dualities}",
    eprint = "2211.13047",
    archivePrefix = "arXiv",
    primaryClass = "hep-th",
    doi = "10.1007/JHEP05(2023)017",
    journal = "JHEP",
    volume = "05",
    pages = "017",
    year = "2023"
}

@article{Chen:2024ddr,
    author = "Chen, Jing-Yuan",
    title = "{Instanton density operator in lattice QCD from higher category theory}",
    eprint = "2406.06673",
    archivePrefix = "arXiv",
    primaryClass = "hep-lat",
    doi = "10.21468/SciPostPhys.19.6.158",
    journal = "SciPost Phys.",
    volume = "19",
    number = "6",
    pages = "158",
    year = "2025"
}

@article{Luscher:1981zq,
    author = "L{\"u}scher, M.",
    title = "{Topology of Lattice Gauge Fields}",
    reportNumber = "BUTP-10/1981-BERN",
    doi = "10.1007/BF02029132",
    journal = "Commun. Math. Phys.",
    volume = "85",
    pages = "39",
    year = "1982"
}

@article{Choi:2021kmx,
    author = "Choi, Yichul and Cordova, Clay and Hsin, Po-Shen and Lam, Ho Tat and Shao, Shu-Heng",
    title = "{Noninvertible duality defects in 3+1 dimensions}",
    eprint = "2111.01139",
    archivePrefix = "arXiv",
    primaryClass = "hep-th",
    reportNumber = "MIT-CTP/5359",
    doi = "10.1103/PhysRevD.105.125016",
    journal = "Phys. Rev. D",
    volume = "105",
    number = "12",
    pages = "125016",
    year = "2022"
}

@article{Cheng:2022sgb,
    author = "Cheng, Meng and Seiberg, Nathan",
    title = "{Lieb-Schultz-Mattis, Luttinger, and 't Hooft - anomaly matching in lattice systems}",
    eprint = "2211.12543",
    archivePrefix = "arXiv",
    primaryClass = "cond-mat.str-el",
    doi = "10.21468/SciPostPhys.15.2.051",
    journal = "SciPost Phys.",
    volume = "15",
    number = "2",
    pages = "051",
    year = "2023"
}

@article{Berkowitz:2023pnz,
    author = "Berkowitz, Evan and Cherman, Aleksey and Jacobson, Theodore",
    title = "{Exact lattice chiral symmetry in 2D gauge theory}",
    eprint = "2310.17539",
    archivePrefix = "arXiv",
    primaryClass = "hep-lat",
    doi = "10.1103/PhysRevD.110.014510",
    journal = "Phys. Rev. D",
    volume = "110",
    number = "1",
    pages = "014510",
    year = "2024"
}

@article{Aoki:2026pvq,
    author = "Aoki, Shoto and Kikukawa, Yoshio and Takemoto, Toshinari",
    title = "{Exact SL(2,Z)-Structure of Lattice Maxwell Theory with $θ$-term in Modified Villain Formulation}",
    eprint = "2604.08736",
    archivePrefix = "arXiv",
    primaryClass = "hep-lat",
    reportNumber = "RIKEN-iTHEMS-Report-26, UT-Komaba/26-4",
    month = "4",
    year = "2026"
}

@article{Seifnashri:2026ema,
    author = "Seifnashri, Sahand",
    title = "{Exactly Solvable 1+1d Chiral Lattice Gauge Theories}",
    eprint = "2601.14359",
    archivePrefix = "arXiv",
    primaryClass = "hep-th",
    month = "1",
    year = "2026"
}

@article{Aoki:2023lqp,
    author = "Aoki, Shoto and Fukaya, Hidenori and Kan, Naoto and Koshino, Mikito and Matsuki, Yoshiyuki",
    title = "{Magnetic monopole becomes dyon in topological insulators}",
    eprint = "2304.13954",
    archivePrefix = "arXiv",
    primaryClass = "cond-mat.mes-hall",
    reportNumber = "OU-HET-1181",
    doi = "10.1103/PhysRevB.108.155104",
    journal = "Phys. Rev. B",
    volume = "108",
    number = "15",
    pages = "155104",
    year = "2023"
}

@article{Slagle:2017wrc,
    author = "Slagle, Kevin and Kim, Yong Baek",
    title = "{Quantum Field Theory of X-Cube Fracton Topological Order and Robust Degeneracy from Geometry}",
    eprint = "1708.04619",
    archivePrefix = "arXiv",
    primaryClass = "cond-mat.str-el",
    doi = "10.1103/PhysRevB.96.195139",
    journal = "Phys. Rev. B",
    volume = "96",
    number = "19",
    pages = "195139",
    year = "2017"
}

@article{Pretko:2018jbi,
    author = "Pretko, Michael",
    title = "{The Fracton Gauge Principle}",
    eprint = "1807.11479",
    archivePrefix = "arXiv",
    primaryClass = "cond-mat.str-el",
    doi = "10.1103/PhysRevB.98.115134",
    journal = "Phys. Rev. B",
    volume = "98",
    number = "11",
    pages = "115134",
    year = "2018"
}

@article{Slagle:2018swq,
    author = "Slagle, Kevin and Aasen, David and Williamson, Dominic",
    title = "{Foliated Field Theory and String-Membrane-Net Condensation Picture of Fracton Order}",
    eprint = "1812.01613",
    archivePrefix = "arXiv",
    primaryClass = "cond-mat.str-el",
    doi = "10.21468/SciPostPhys.6.4.043",
    journal = "SciPost Phys.",
    volume = "6",
    number = "4",
    pages = "043",
    year = "2019"
}

@article{You:2019ciz,
    author = "You, Yizhi and Devakul, Trithep and Sondhi, S. L. and Burnell, F. J.",
    title = "{Fractonic Chern-Simons and BF theories}",
    eprint = "1904.11530",
    archivePrefix = "arXiv",
    primaryClass = "cond-mat.str-el",
    doi = "10.1103/PhysRevResearch.2.023249",
    journal = "Phys. Rev. Res.",
    volume = "2",
    number = "2",
    pages = "023249",
    year = "2020"
}

@article{Gorantla:2020xap,
    author = "Gorantla, Pranay and Lam, Ho Tat and Seiberg, Nathan and Shao, Shu-Heng",
    title = "{More Exotic Field Theories in 3+1 Dimensions}",
    eprint = "2007.04904",
    archivePrefix = "arXiv",
    primaryClass = "cond-mat.str-el",
    doi = "10.21468/SciPostPhys.9.5.073",
    journal = "SciPost Phys.",
    volume = "9",
    pages = "073",
    year = "2020"
}

@article{Slagle:2020ugk,
    author = "Slagle, Kevin",
    title = "{Foliated Quantum Field Theory of Fracton Order}",
    eprint = "2008.03852",
    archivePrefix = "arXiv",
    primaryClass = "hep-th",
    doi = "10.1103/PhysRevLett.126.101603",
    journal = "Phys. Rev. Lett.",
    volume = "126",
    number = "10",
    pages = "101603",
    year = "2021"
}

@article{Gorantla:2020jpy,
    author = "Gorantla, Pranay and Lam, Ho Tat and Seiberg, Nathan and Shao, Shu-Heng",
    title = "{fcc lattice, checkerboards, fractons, and quantum field theory}",
    eprint = "2010.16414",
    archivePrefix = "arXiv",
    primaryClass = "cond-mat.str-el",
    doi = "10.1103/PhysRevB.103.205116",
    journal = "Phys. Rev. B",
    volume = "103",
    number = "20",
    pages = "205116",
    year = "2021"
}

@article{Yamaguchi:2021qrx,
    author = "Yamaguchi, Satoshi",
    title = "{Supersymmetric quantum field theory with exotic symmetry in 3+1 dimensions and fermionic fracton phases}",
    eprint = "2102.04768",
    archivePrefix = "arXiv",
    primaryClass = "hep-th",
    reportNumber = "OU-HET 1090",
    doi = "10.1093/ptep/ptab037",
    journal = "PTEP",
    volume = "2021",
    number = "6",
    pages = "063B04",
    year = "2021"
}

@article{Hsin:2021mjn,
    author = "Hsin, Po-Shen and Slagle, Kevin",
    title = "{Comments on foliated gauge theories and dualities in 3+1d}",
    eprint = "2105.09363",
    archivePrefix = "arXiv",
    primaryClass = "cond-mat.str-el",
    reportNumber = "CALT-TH-2021-022",
    doi = "10.21468/SciPostPhys.11.2.032",
    journal = "SciPost Phys.",
    volume = "11",
    number = "2",
    pages = "032",
    year = "2021"
}

@article{Geng:2021cmq,
    author = "Geng, Hao and Kachru, Shamit and Karch, Andreas and Nally, Richard and Rayhaun, Brandon C.",
    title = "{Fractons and Exotic Symmetries from Branes}",
    eprint = "2108.08322",
    archivePrefix = "arXiv",
    primaryClass = "hep-th",
    doi = "10.1002/prop.202100133",
    journal = "Fortsch. Phys.",
    volume = "69",
    number = "11-12",
    pages = "2100133",
    year = "2021"
}

@article{Yamaguchi:2021xeq,
    author = "Yamaguchi, Satoshi",
    title = "{Gapless edge modes in (4+1)-dimensional topologically massive tensor gauge theory and anomaly inflow for subsystem symmetry}",
    eprint = "2110.12861",
    archivePrefix = "arXiv",
    primaryClass = "hep-th",
    reportNumber = "OU-HET 1111",
    doi = "10.1093/ptep/ptac032",
    journal = "PTEP",
    volume = "2022",
    number = "3",
    pages = "033B08",
    year = "2022"
}

@article{Gorantla:2022eem,
    author = "Gorantla, Pranay and Lam, Ho Tat and Seiberg, Nathan and Shao, Shu-Heng",
    title = "{Global dipole symmetry, compact Lifshitz theory, tensor gauge theory, and fractons}",
    eprint = "2201.10589",
    archivePrefix = "arXiv",
    primaryClass = "cond-mat.str-el",
    doi = "10.1103/PhysRevB.106.045112",
    journal = "Phys. Rev. B",
    volume = "106",
    number = "4",
    pages = "045112",
    year = "2022"
}

@article{Gorantla:2022ssr,
    author = "Gorantla, Pranay and Lam, Ho Tat and Seiberg, Nathan and Shao, Shu-Heng",
    title = "{(2+1)-dimensional compact Lifshitz theory, tensor gauge theory, and fractons}",
    eprint = "2209.10030",
    archivePrefix = "arXiv",
    primaryClass = "cond-mat.str-el",
    reportNumber = "MIT-CTP/5462, YITP-SB-2022-31",
    doi = "10.1103/PhysRevB.108.075106",
    journal = "Phys. Rev. B",
    volume = "108",
    number = "7",
    pages = "075106",
    year = "2023"
}

@article{Ohmori:2022rzz,
    author = "Ohmori, Kantaro and Shimamura, Shutaro",
    title = "{Foliated-exotic duality in fractonic BF theories}",
    eprint = "2210.11001",
    archivePrefix = "arXiv",
    primaryClass = "hep-th",
    doi = "10.21468/SciPostPhys.14.6.164",
    journal = "SciPost Phys.",
    volume = "14",
    number = "6",
    pages = "164",
    year = "2023"
}

@article{Honda:2022shd,
    author = "Honda, Masazumi and Nakanishi, Taiichi",
    title = "{Scalar, fermionic and supersymmetric field theories with subsystem symmetries in d + 1 dimensions}",
    eprint = "2212.13006",
    archivePrefix = "arXiv",
    primaryClass = "hep-th",
    reportNumber = "YITP-22-135, RIKEN-iTHEMS-Report-22",
    doi = "10.1007/JHEP03(2023)188",
    journal = "JHEP",
    volume = "03",
    pages = "188",
    year = "2023"
}

@article{Hsin:2023ooo,
    author = "Hsin, Po-Shen and Luo, Zhu-Xi and Malladi, Ananth",
    title = "{Gapped interfaces in fracton models and foliated fields}",
    eprint = "2308.04489",
    archivePrefix = "arXiv",
    primaryClass = "cond-mat.str-el",
    doi = "10.1007/JHEP11(2023)089",
    journal = "JHEP",
    volume = "11",
    pages = "089",
    year = "2023"
}

@article{Shimamura:2024kwf,
    author = "Shimamura, Shutaro",
    title = "{Anomaly of subsystem symmetries in exotic and foliated BF theories}",
    eprint = "2404.10601",
    archivePrefix = "arXiv",
    primaryClass = "cond-mat.str-el",
    doi = "10.1007/JHEP06(2024)002",
    journal = "JHEP",
    volume = "06",
    pages = "002",
    year = "2024"
}

@article{Ebisu:2024mbb,
    author = "Ebisu, Hiromi and Honda, Masazumi and Nakanishi, Taiichi and Shimamori, Soichiro",
    title = "{New field theories with foliation structure and subdimensional particles from the Godbillon-Vey invariant}",
    eprint = "2408.05048",
    archivePrefix = "arXiv",
    primaryClass = "hep-th",
    reportNumber = "YITP-24-78, RIKEN-iTHEMS-Report-24, OU-HET-1237",
    doi = "10.1103/4fz5-9298",
    journal = "Phys. Rev. D",
    volume = "112",
    number = "2",
    pages = "025010",
    year = "2025"
}

@article{Witten:1979ey,
    author = "Witten, Edward",
    title = "{Dyons of Charge e theta/2 pi}",
    reportNumber = "CERN-TH-2724",
    doi = "10.1016/0370-2693(79)90838-4",
    journal = "Phys. Lett. B",
    volume = "86",
    pages = "283--287",
    year = "1979"
}

@article{Cao:2023doz,
    author = "Cao, Weiguang and Li, Linhao and Yamazaki, Masahito and Zheng, Yunqin",
    title = "{Subsystem non-invertible symmetry operators and defects}",
    eprint = "2304.09886",
    archivePrefix = "arXiv",
    primaryClass = "cond-mat.str-el",
    doi = "10.21468/SciPostPhys.15.4.155",
    journal = "SciPost Phys.",
    volume = "15",
    number = "4",
    pages = "155",
    year = "2023"
}

\end{document}